# Diurnal temperature variations and migrating thermal tides in the Martian lower atmosphere observed by the Emirates Mars InfraRed Spectrometer

Siteng Fan[1,2], François Forget[2], Michael D. Smith[3], R. John Wilson[4], Sandrine Guerlet[2], Khalid M. Badri[5], Samuel A. Atwood[3,6], Roland M. B. Young[7,8], Christopher S. Edwards[9], Philip R. Christensen[10], Justin Deighan[11], Hessa R. Al Matroushi[5], Antoine Bierjon[2], Jiandong Liu[2], Ehouarn Millour[2]

[1]Department of Earth and Space Sciences, Southern University of Science and Technology, Shenzhen, China
[2]LMD/IPSL, Sorbonne Université, PSL Research Université, École Normale Supérieure, École Polytechnique, CNRS, Paris, France
[3]NASA Goddard Space Flight Center, Greenbelt, MD, USA
[4]NASA Ames Research Center, Mountain View, CA, USA
[5]Mohammed Bin Rashid Space Centre, Dubai, UAE
[6]University of Maryland Baltimore County, Baltimore, MD, USA
[7]Department of Physics, SUPA, University of Aberdeen, King's College, Aberdeen, UK
[8]Department of Physics & National Space Science and Technology Center, United Arab Emirates University, Al Ain, UAE
[9]Department of Astronomy and Planetary Science, Northern Arizona University, Flagstaff, AZ, USA
[10]School of Earth and Space Exploration, Arizona State University, Tempe, AZ, USA
[11]Laboratory for Atmospheric and Space Physics, University of Colorado Boulder, Boulder, CO, USA

Corresponding author: Siteng Fan (fanst@sustech.edu.cn)

**Key Points:**
- Diurnal temperature variations in the Martian atmosphere are observed at all local times on a planetary scale for an entire Martian year.
- The updated Mars Planetary Climate Model shows good agreement with observations, but has differences in tide amplitudes and structures.
- Zonal winds and dust content regulate the seasonal variations of the migrating thermal tides.

**Abstract**
The Martian atmosphere experiences large diurnal variations due to the ~24.6 h planetary rotation and its low heat capacity. Understanding such variations on a planetary scale is limited due to the lack of observations, which are greatly addressed with the recent advent of the Emirates Mars Mission (EMM). As a result of its unique high-altitude orbit, instruments onboard are capable of obtaining a full geographic and local time coverage of the Martian atmosphere every 9-10 Martian days, approximately ~5° in solar longitude ($L_S$). This enables investigations of the diurnal variation of the current climate on Mars on a planetary scale without significant local time (LT) gaps or confusions from correlated seasonal variations. Here, we present the results of diurnal temperature variations and thermal tides in the Martian atmosphere using temperature profiles retrieved from the Emirates Mars InfraRed Spectrometer (EMIRS) observations. The data during the primary mission is included, covering an entire Martian Year (MY) starting from MY 36 $L_S = 49°$. The diurnal temperature patterns suggest a dominant diurnal tide in most seasons, while the semi-diurnal tide presents a similar amplitude near perihelion. The seasonal variation of the diurnal tide latitudinal distribution is well explained by the total vorticity due to zonal wind, while that of the semi-diurnal tide following both dust and water ice clouds, and the ter-diurnal tide following the dust only. Comparison with the updated Mars Planetary Climate Model (PCM, version 6) suggest improvements in simulating the dust and water cycles, as well as their radiative processes.

**Plain language summary**
As a result of Mars' planetary rotation of ~24.6 h and the low mass of its atmosphere, the atmospheric temperature changes dramatically every day following the cycle of the incoming sunlight. Such rapid and large changes, together with dust and clouds, can excite a number of waves propagating in the Martian atmosphere. These strong waves are one of the key factors controlling the weather and climate on Mars. Although these temperature variations and waves are important, our understanding is currently limited due to the lack of observations. A recent spacecraft mission located in an orbit distant from Mars has largely solved this problem by providing observations with good coverage in both location and time, enabling the investigations on the temperature variations and waves on a global and hourly scale. Here, we present the results of such investigations during the first Martian year of this mission, as well as comparison with the outputs from a Mars climate model, which represents our current knowledge. The model and the observations show generally good agreements, with differences suggesting directions in future model improvements, among which are the location and time of the dust and clouds in the Martian atmosphere.

## 1. Introduction

The Martian atmosphere is an ideal independent natural laboratory to study terrestrial planetary atmospheres. Similar to those on the Earth, physical and dynamical processes in the thin atmosphere of Mars are primarily driven by solar insolation. As a result of the fast planetary rotation with a period of ~24.6 h (Proctor, 1873) and small atmosphere and surface heat capacities (Gierasch & Goody, 1968; Morrison et al., 1969), the Martian atmosphere experiences large diurnal variations, e.g., ranges of 100 Pa and 80 K in surface pressure and temperature, respectively (Banfield et al., 2020). Strong atmospheric waves are therefore excited in the atmosphere (Gierasch & Goody 1968; Lindzen & Chapman, 1969), where thermal tides play a dominant role.

Thermal tides are zonally and vertically propagating harmonic responses to the diurnal solar forcing, and are further regulated by planetary topography (Zurek, 1976), atmospheric radiation sources (e.g., dust and clouds; Wilson & Hamilton, 1996; Zurek & Leovy, 1981), and general circulation (Takahashi et al., 2006). These tides are planetary-scale with integer zonal wavenumbers and periods of integer fractions of a Martian solar day (sol) (Lindzen & Chapman, 1969), typically present in the fields of temperature, wind, surface pressure, air density, and composition, etc. (Hess et al., 1977; Banfield et al., 2003; 2020; Lee et al., 2009; Withers et al., 2011; Guzewich et al., 2016; Sánchez-Lavega et al., 2023). In the lower atmosphere, the tides usually have large amplitudes at all times (e.g., ~10 K near aphelion and much larger in other seasons; Lee et al., 2009), and some modes significantly increase while propagating upward to altitudes with low air densities as a result of the conservation of energy (Lindzen & Chapman, 1969). These tides are among the dominating factors controlling diurnal physical and dynamical processes in the Martian atmosphere, and their propagation also connects different atmospheric layers (Forbes, 1995). Despite these importances, characterizing thermal tides in the Martian atmosphere and understanding their related physical and dynamical processes face difficulties due to the coupling among tides, dust, clouds, and general circulation (Wu et al., 2022). On one hand, the air-lifted dust, clouds, and the atmospheric dynamics excite/regulate the tides and influence their propagation (Hamilton, 1982; Wilson & Hamilton, 1996; Zurek & Leovy, 1981; Takahashi et al., 2006; Wilson et al., 2014; Wu et al., 2015; 2017). On the other hand, the tides affect the transport of dust and clouds, dust lifting, and cloud formation (Forbes 1995; Spiga et al., 2016; Hinson & Wilson, 2004; Wilson & Guzewich, 2014). They also deposit momentum to the general circulation (Moudden & Forbes, 2008; Toigo et al., 2012).

Diurnal temperature variations in the Martian atmosphere were first observed by the Michelson interferometer on Mariner 9 (Hanel et al., 1972a; 1972b), and followed by the Viking InfraRed Thermal Mapper (IRTM; Kieffer et al., 1972), where thermal tides accounts for ~15 K contrast between the surface and ~500 Pa atmosphere represented by the 15 μm brightness temperature (Wilson & Richardson, 2000). Such efforts were continued using temperature profiles obtained by the Thermal Emission Spectrometer onboard the Mars Global Surveyor (MGS/TES; Conrath et al., 2000). A number of atmosphere wave modes were resolved with vertical resolutions (Banfield et al., 2000; 2003; 2004; Wilson, 2000). Observations from the Mars Climate Sounder onboard Mars Reconnaissance Orbiter (MRO/MCS; McCleese et al., 2007) were then used to obtain comprehensive climatology of the Martian atmosphere with thermal tides extensively analyzed (Lee et al., 2009; Kleinböhl et al., 2013; Wu et al., 2015; 2017; 2021). However, given that these two Martian orbiters were in Sun-synchronous orbits, their temperature observations only sampled at local times around 2 am/pm and 3 am/pm, respectively. Such limited local time sampling introduces strong aliasing in the analyses of atmospheric wave modes, particularly for the semi-diurnal tide. Although cross-track limb sounding was performed to partially relieve this aliasing (Kleinböhl et al., 2013), these observations are unevenly-spaced in local time and large uncertainties still exist. Trying to address this issue, some later temperature observations were obtained from drifting orbits, e.g., the Planetary Fourier Spectrometer onboard the Mars Express (MEX/PFS; Giuranna et al., 2021) and the TIRVIM Fourier-spectrometer, part of the Atmospheric Chemistry Suite onboard the ExoMars Trace Gas Orbiter (TGO/ACS/TIRVIM; Guerlet et al., 2022), but the slow drifting results in contaminations of seasonal changes in the diurnal analyses. PFS requires ~1 Martian year (MY) to complete a full coverage of local time (Giuranna et al., 2021), which prevents distinguishing diurnal variations from seasonal changes. The sub-spacecraft local time drifting of TGO is faster, but it still needs ~55 Martian solar days (sols) to obtain full local time coverage of nadir observations. The TIRVIM instrument onboard samples more frequently around 9 am/pm, complementing those of MCS (Guerlet et al., 2022) and making it possible to analyze the planetary-scale sub-diurnal thermal tides (Fan et al., 2022a). However, significant influence from seasonal variations still exists. The sampling drifting in local time leads to overestimates of some wave modes with frequencies close to the sampling (Fan et al., 2022a). Further corrections, e.g., detrending the seasonal variations, are necessary when the Martian atmosphere rapidly changes (Guerlet et al., 2023). To fully address such issues, a recent spacecraft in a novelly high-altitude orbit, the Hope probe of the Emirates Mars Mission (EMM; Amiri et al., 2021; Almatroushi et al., 2021), provides observations with broad coverage in both location and local time, which enables detailed analyses of diurnal climate on Mars. The temperature profiles (Smith et al., 2022) retrieved from the Emirates Mars InfraRed Spectrometer (EMIRS; Edwards et al., 2021) observations are ideal for the analysis of diurnal temperature variations and related thermal tides, which is the subject of this work.

The rest of the paper is arranged as follows. Section 2 overviews the spacecraft, Hope probe, and the instrument, EMIRS, as well as the Mars Planetary Climate Model (PCM) that serves as a benchmark of our current knowledge. Section 3 details the methodology, including data processing and wave mode analysis. Section 4 presents the results of the diurnal mean temperatures, diurnal temperature anomalies, and amplitudes and phases of migrating thermal tides in the two solstitial and the two equinoctial seasons. Seasonal variations of the tide amplitudes are also included. Section 5 discusses the findings and compares them with previous studies, followed by a conclusion in Section 6.

## 2. Observation and simulation
### 2.1 EMM/EMIRS
The EMM Hope probe (Amiri et al., 2021; Almatroushi et al., 2021) is the first mission to Mars led by the United Arab Emirates (UAE). It was launched in July 2020, and started its science phase on May 23$^{rd}$, 2021 (MY 36 $L_S$ = 49°). Given its three scientific objectives focusing on the Martian atmosphere with emphases on the planetary-scale diurnal/sub-diurnal investigations, the spacecraft was designed to be located in a lowly inclined (~25°) high-altitude orbit (19,970/42,650 km altitude at periapsis/apoapsis) with an orbital period of 54.5 hr. As a result, the three remote sensing instruments onboard (Jones et al., 2021; Edwards et al., 2021; Holsclaw et al., 2021) are capable of observing the entire disk of Mars from any point in orbit at wavelengths from ultraviolet to infrared. Observing while orbiting the rotating Mars, these instruments can achieve nearly complete coverage of all geographic locations and local times within 9-10 sols, which facilitates comprehensive studies of the Martian atmosphere on a planetary and diurnal scale. In addition, studies of interactions from the lower to the upper atmospheres are possible through observation synergy among the three instruments.

EMIRS (Edwards et al., 2021) is a Fourier transform infrared spectrometer. It covers a wavelength range of 6-100 μm (1666-100 cm$^{-1}$) with a selectable spectral resolution of 5 or 10 cm$^{-1}$. The spectral range covered by EMIRS includes the features used for the retrieval of atmospheric temperature ($CO_2$ band at 15 μm), surface temperature (continuum at 7-12 μm), column integrated dust (9 μm), water ice (12 μm), and water vapor (25-40 μm). The instrument is equipped with a moving pointing mirror in the optical path to scan the Martian disk with a footprint size of 100 km (near periapsis) to 300 km (near apoapsis). It takes ~0.5 hr for each scanning sequence covering the entire spacecraft-facing Martian disk, which is typically scheduled ~20 times per orbit, making it possible for the aforementioned diurnal variation investigations.

2.2 Temperature retrieval

The temperature profiles are part of the Level 3 (L3) product of the EMIRS data available at the EMM Science Data Center (https://sdc.emiratesmarsmission.ae/). The retrieval procedure of EMIRS infrared spectra (Smith et al., 2022) follows the inversion algorithm derived for TES (Conrath et al, 2000; Smith et al., 2006). It retrieves the surface temperature, the atmospheric temperature from surface to ~40 km (~4 scale heights), the column densities of dust, water ice clouds, and water vapor. As the spectral features of these quantities are well-separated in wavelength, the algorithm iteratively optimizes (1) the surface and atmospheric temperatures, (2) column optical depths of dust and water ice clouds, and (3) the column density of water vapor. The retrieval algorithm takes the surface pressure from the Mars Climate Database (MCD; Forget et al., 1999; Millour et al., 2018), as it can not be reliably constrained from observations. This may introduce systematic biases, which, however, are expected to be small as the MCD has been well validated against a number of surface pressure observations, including the two Viking landers, InSight, and Perseverance (Millour et al., 2024). The dust and water ice cloud particles sizes are assumed to be 1.5 μm and 4.0 μm, consistent with the latest results (Atwood et al., 2024). The vertical distribution of dust is assumed to follow a Conrath profile (Conrath, 1975), and that of the water ice clouds are assumed to be all located at the condensation level above an atmosphere with well-mixed water vapor up to this level. A recent update of the algorithm (Smith et al., 2025) improved the retrieved diurnal cycles of dust and water vapor by adjusting their effective temperatures, but it does not influence the derived temperature profiles constrained from the $CO_2$ band (see below). Spherical geometry is neglected in the retrieval, which is only applied to observations with low emission angles (<70°).

The temperature profiles retrieved using the EMIRS spectra are constrained by fitting the depth and shape of the $CO_2$ 15 μm band (667 cm$^{-1}$) with a number of assumptions (Smith et al., 2022). The vertical smoothness is assumed to follow that in Conrath et al., (2000) with a correlation length of 0.75 scale heights because of the inherent width of the $CO_2$ contribution functions. The temperature profiles at the bottom 0.75 scale heights are assumed to have the same lapse rates as those in the MCD, and temperatures above 4 scale heights are the same as those extracted from the MCD. Therefore, the retrieved temperature at each reported height is not independent. Instead, they are oversampled compared to their equivalent resolution of 1-1.5 scale heights. Vertical averaging kernels are correspondingly derived for each temperature profile. Uncertainties of the derived temperature measurements are dominated by systematic errors in the retrieval procedures (Smith et al., 2022). They are estimated to be 2 K at 1-3 scale heights, which increases to 5 K at 0.5-1 and 3-3.5 scale heights, and to 10 K at 0.25-0.5 and 3.5-4 K scale heights.

2.3 Mars PCM

The Mars PCM (Forget et al., 1999), previously known as the Laboratoire de Météorologie Dynamique Mars Global Climate Model (LMD Mars GCM), serves as a benchmark of our current understanding of the Martian climate. It is a general circulation model based on the LMDZ dynamical core, which uses a latitude-longitude C-grid with a shallow-atmosphere hydrostatic assumption. Forcings are computed in each vertical column through one-dimensional physical processes, including (1) the radiative effects of $CO_2$, dust, and clouds, (2) the $CO_2$, dust, and water cycles with cloud microphysics, (3) the transports of gases and dust, (4) sub-grid surface ice interactions, (5) photochemistry and isotope cycles, etc. The model has been continuously upgraded in the recent decades (e.g., Lefèvre et al., 2008; Madeleine et al. 2011; Colaïtis et al., 2013; González-Galindo et al., 2013; 2015; Navarro et al. 2014; Forget et al., 2022). A number of new/updated processes have been implemented since the last version (version 5), including dust injection schemes of mountain top flows and rocket dust storms (Wang et al., 2018), non-orographic gravity waves (Gilli et al., 2020; Liu et al., 2023; 2025), the HDO cycle (Vals et al., 2022), the $CO_2$ scavenging process (Määttänen et al., 2022), and the surface slope microclimate (Lange et al., 2023), etc.

The simulation results (Fan, 2024) used for model-observation comparison in this work are based on the Mars PCM version 6. The model has a 64×48×73 grid in longitude, latitude, and pressure levels, which corresponds to a horizontal resolution of 5.625° and 3.75° in longitude and latitude, respectively, and 73 hybrid terrain-following vertical pressure levels (σ-grid) from surface to ~2×10$^{-3}$ Pa, including the Martian thermosphere. The run has 960 dynamical timesteps in each Martian day, updating the circulation and transport every 1.5 Martian minutes. The physical timestep is 7.5 Martian minutes, five times that of the dynamics, suggested by the optimization of the water cycle. All the aforementioned physical processes in the version 6 are turned on in the run used in this work, where the dust injection is semi-interactive regulated by the MY 36 and MY 37 dust scenarios built following Montabone et al., (2015; 2020).

### 3. Methodology
3.1 Data processing
The EMIRS data analyzed in this work were obtained during the EMM primary mission from May 23$^{rd}$, 2021 to April 10$^{th}$, 2023, covering an entire Martian year from MY 36 $L_S$ = 49° to MY 37 $L_S$ = 49°. The successful temperature retrievals total ~1.32×10$^6$ in number during this period. They are mostly uniformly distributed in $L_S$ except for some data gaps (Figure 1a) due to a solar conjunction (MY 36 $L_S$ = 100° – 120°) and several spacecraft safe mode events. Given the fact that a full geographic and local time coverage can be achieved within ~10 sols, the data are binned every 5° in $L_S$. The observations within each $L_S$ bin are then assumed to take place on the same sol in the following analyses, where seasonal variations are neglected as this 5° $L_S$ range is sufficiently small.

The retrievals used for producing the EMIRS L3 data are conducted on the same σ-grid, which has the same prescribed ratios of grid pressure levels to the surface pressure. As the surface pressure is different among observations, the derived temperature profiles are interpolated onto the standard pressure grid, so are the corresponding uncertainties and averaging kernels. Such an interpolation may introduce systematic bias at high pressure levels due to large topographic variations (e.g., the Tharsis region), but its influence on the diurnal variations, which are mainly driven by the diurnal solar forcing, are expected to be limited. The interpolated profiles are then binned in latitude, longitude, and local time with bin sizes of 10°, 5°, and 1 hr, respectively, to derive the zonal and diurnal mean and anomaly (Section 4.1). Each bin is assigned the same weight to reduce zonal and local time sampling biases. The uncertainties are computed through error propagation, which are usually negligible due to the large number of observations, and therefore not presented in the figures to avoid possible distractions.

## 3.2 Wave mode decomposition

The temperature profiles interpolated onto the standard pressure grid are further decomposed to analyze the wave modes in the Martian atmosphere. As the thermal tides are harmonic oscillations propagating along parallels of latitude with integer zonal wavenumbers and with temporal frequencies of integer multiples of a sol, the perturbations in the temperature field can be assumed as linear combinations of different wave modes (Gierasch & Goody, 1968; Zurek, 1976). Therefore, observations within each $L_S$ bin are assumed to be an eq-sol diurnal composite in the decomposition, as all these waves should start with the same respective phases on every sol. The decomposition results presented for each 5° $L_S$ bin are analyzed using observations within a 15° $L_S$ range to improve the statistics, including the 5° $L_S$ bins immediately before and after, which leads to 10° $L_S$ overlaps between each pair of two consecutive bins. A test only using data within each bin is presented and discussed in Section 5.

In line with previous analyses using MCS (Wu et al., 2015; 2017), TIRVIM (Fan et al., 2022a), and EMIRS (Fan et al., 2022b) data, least-square fits with prescribed wave modes are applied on a universal time (UT) coordinate in the wave mode decomposition. Under the linear assumption, each observed temperature ($T$) is an arithmetic sum of simple harmonic waves:

$$T(\lambda, \varphi, p, t) = \sum_{\sigma,s}\left(C_{\sigma,s}(\varphi,p) \times \cos(s\lambda + \sigma t) + S_{\sigma,s}(\varphi,p) \times \sin(s\lambda + \sigma t)\right) \quad (1)$$

where $\lambda$, $\varphi$, $p$, and $t$ are the longitude, latitude, pressure level, and UT, respectively; $s$ and $\sigma$ are the frequencies in longitude and time; $C$ and $S$ are the coefficients of the Fourier bases of the waves. The UT coordinate, instead of an LT one, is used in the decomposition, as the observations are near-uniformly distributed in both location and time (Figure 1c-1f). This was an issue in analyzing TES and MCS observations due to their unequal temporal samplings centered around two local times (Wilson, 2000; Banfield et al., 2003; Lee et al., 2009; Kleinböhl et al., 2013). The pairs of integers, ($s$, $\sigma$), denote wave modes, e.g., ($s$, $\sigma$) = (1, 1) is the mode that has a unit zonal wavenumber and a frequency of one per sol, which is the migrating diurnal tide that propagates westwards Sun-synchronously. Using the prevailing nomenclatures used in the tidal theory, "DW1" is used for the notation of this tide mode, where "D" stands for "temporal period of diurnal" ($\sigma = 1$), "W" for "westward propagating" ($s/\sigma > 0$), and "1" for "zonal wavenumber 1" ($s = 1$). Therefore, the semi- and ter-diurnal tides, which are also Sun-synchronous migrating tides, are denoted as SW2 ($s = \sigma = 2$) and TW3 ($s = \sigma = 3$), respectively. $C$ and $S$ are the unknowns to fit in the analysis. The amplitude ($A$) and phase ($\theta$) of each wave mode, defined by the ($s$, $\sigma$) pair, have the following relationships with these coefficients.

$$A_{\sigma,s} = \sqrt{C_{\sigma,s}^2 + S_{\sigma,s}^2} \quad (2)$$

$$\theta_{\sigma,s} = \tan^{-1}\left(\frac{C_{\sigma,s}}{S_{\sigma,s}}\right) \quad (3)$$

Temperature measurements within each $L_S$ and latitude bin, whose sizes are 15° and 10°, respectively, are considered individually in the decomposition. At each pressure level, Equation (1) for all observations within this bin can be combined and rearranged to the following form for linear regression.

$$\mathbf{W}_{[N \times D]} \mathbf{x}_{[D \times 1]} + \mathbf{e}_{[N \times 1]} = \mathbf{y}_{[N \times 1]} \quad (4)$$

where $\mathbf{W}$ is the weight matrix with its ($i$, $j$)-th element being the value of the $j$-th Fourier basis,

either that of the sine or cosine function in Equation (1), of the $i$-th observation; $\mathbf{x}$ is the coefficient vector to fit, consisting of the unknown Fourier coefficients, $C$ and $S$; $\mathbf{y}$ and $\mathbf{e}$ are the observation vectors containing the temperature measurements and their corresponding uncertainties, respectively; $N$ is the number of observations, which usually has the value of a few thousands at low- to mid-latitudes (Figure 1g-1j); $D$ is the degree of freedom, which is the total number of linearly independent coefficients. The wave mode truncation is set as $s = \{0, 1, 2, 3\}$ and $\sigma = \{-3, -2, -1, 0, 1, 2, 3\}$, which leads to $D = 49$. Such a truncation is proper to extract all major wave modes, suggested by the residuals below the observation uncertainties. The TW3 mode is included in all analyses in this work, which is shown distinguishable and necessary during the aphelion season (Fan et al., 2022b), and are likely more prominent during other seasons.

Finally, the values of the Fourier coefficients, the $\mathbf{x}$ vector, can be solved by applying a least-square fit to the linear system with the expected values ($\mathbf{E}[\mathbf{x}]$) and covariance ($\mathbf{Cov}[\mathbf{x}]$) as follows.

$$\mathbf{E}[x] = (\mathbf{W}^T\mathbf{C}^{-1}\mathbf{W})^{-1}(\mathbf{W}^T\mathbf{C}^{-1})\mathbf{y} \qquad (5)$$

$$\mathbf{Cov}[x] = [(\mathbf{W}^T\mathbf{C}^{-1}\mathbf{W})^{-1}(\mathbf{W}^T\mathbf{C}^{-1})e][(\mathbf{W}^T\mathbf{C}^{-1}\mathbf{W})^{-1}(\mathbf{W}^T\mathbf{C}^{-1})e]^T$$
$$= (\mathbf{W}^T\mathbf{C}^{-1}\mathbf{W})^{-1} \qquad (6)$$

where $\mathbf{C}$ is the observation covariance matrix, a diagonal matrix with its $i$-th diagonal element being the square of the $i$-th element of $\mathbf{e}$, assuming the observations are independent. Uncertainties of the fitted coefficients are the square roots of the diagonal elements of $\mathbf{Cov}[\mathbf{x}]$ by definition, and those of the amplitudes and phases can be derived through error propagation using Equations (2) and (3).

## 4. Results
### 4.1 Temperature observation and coverage

As a result of the orbit design of the EMM Hope spacecraft and the EMIRS instrument observation strategy, the temperature observations are mostly uniformly distributed in longitude and local time during all seasons (Figure 1a, 1c-1f), except for some small data gaps in $L_S$ due to spacecraft safe events and a big one around MY 36 $L_S = 100° – 120°$ due to a Mars solar conjunction (Figure 1a). The data obtained within 5° $L_S$ before each solstice and each equinox are selected in the following analyses ($L_S = 85° – 90°$ for summer solstice, $L_S = 175° – 180°$ for autumn equinox, $L_S = 265° – 270°$ for winter solstice, and $L_S = 355° – 360°$ for spring equinox). All of them have full geographic and temporal coverages (Figure 1c-1f). The Martian year covered by the EMM primary mission shows a typical dust annual cycle without a global dust storm (Figure 1b). Two large regional dust storms, the A- and C-storms, labelled in line with Kass et al., (2016), occurred near $L_S = 220°$ and $L_S = 310°$, and two small ones, an unusual one and the B-storm, around $L_S = 155°$ and $L_S = 255°$. The four selected seasons are outside these dust storms with their dust contents similar to other MYs, so the results could represent the climatology on Mars. The global average dust column optical depths ($\tau$) during these four selected $L_S$ (dashed lines in Figure 1b) are all below 0.3 with an increase from $\tau$ <0.1 during dust-free summer to $\tau$ ~0.25 during the dusty winter, which then decreases towards the spring equinox of the next MY.

The zonal total numbers of successful EMIRS retrievals in each (latitude, local time) bin within 5° $L_S$ range show near-uniform distributions and full coverages in local time (Figure 1g-1j), except for fewer nighttime/daytime observations in the summer/winter hemisphere due to the low orbit inclination (Section 2.1). This also results in fewer observations at high latitudes than the equatorial region. There are usually >2000 observations in the equatorial 10° latitude bin, which become ~500 near ±60°. These large zonal total numbers lead to good statistics in deriving the temperature variations and conducting wave mode decompositions. As shown in the binned temperature profiles (color lines in Figure 1k-1n), diurnal and seasonal temperature variations are significant in the Martian atmosphere. The equatorial region is typically ~20 K warmer during the northern winter than summer due to smaller Sun-Mars distance, with an average temperature of 197 K at 100 Pa and 172 K at 10 Pa at $L_S = 265° – 270°$ (Figure 1m), compared to those of 175 K at 100 Pa and 152 K at 10 Pa at $L_S = 85° – 90°$ (Figure 1k). The vertical temperature gradient is also smaller near the winter solstice (Figure 1m), which is <50 K between 500 Pa and 5 Pa compared to ~ 58 K in all the other three seasons (Figure 1k, 1m, 1n). This is likely due to vertically extended airborne dust in the declining phase of a regional dust storm (Figure 1b). Besides differences among seasons, these temperature profiles all show features of thermal tides (Figure 1k-1n), where the diurnal temperature maxima and minima alternate in local time at different pressure levels. Warmer mornings are usually at >100 Pa and <10 Pa, and warmer afternoons in between, mainly due to the strong migrating tides in the Martian atmosphere (see Section 4.3 and 4.4). The vertical smoothing of these profiles (black solid lines in Figure 1k-1n) in the form of retrieval kernels is similar to that of TES observations, approximately half the vertical resolution of those of MCS, which mainly originates from the difference between nadir (TES) and limb (MCS) observing geometries.

4.2 Diurnal mean temperature
Zonal and diurnal mean temperatures (Figure 2a-2d) are computed by averaging the binned temperature along the longitude and the local time axes, which are shown between 500 and 5 Pa, equivalent to a typical altitude range from near surface to ~45 km. The uncertainty of the binned temperature is small, <0.5 K at most pressure levels at <80° latitude, due to the large number of observations. Large topographical anomalies (e.g., the Tharsis region) may influence the pressure and temperature, but they are also part of the "zonal mean". The northern winter (Figure 2c) is globally warmer than the summer (Figure 2a) by ~20 K due to higher total solar flux as well as higher atmospheric dust content. These latitude-pressure temperature distributions present typical seasonal features. During the two solstice seasons (Figure 2a, 2c), the summer polar region is usually tens of degrees warmer than the winter one, and there exists a warming structure over the winter pole due to adiabatic warming of the downwelling branch of the one-cell solstitial Hardley circulation (Forget et al., 1999). The patterns in the two equinox seasons (Figure 2b, 2d) are more north-south symmetric with maxima in the equatorial region and two warming structures above both poles because of the two-cell equinoctial Hardley circulation. The temperature in the southern hemisphere near the northern winter equinox (Figure 2c) is significantly higher than that at all other latitudes or during other seasons, which is likely due to the regional dust storm at southern high latitudes (south of 60° S, Figure 1b).

The Mars PCM (Section 2.3; Forget et al., 1999) results show similar zonal and diurnal mean temperature distributions (Figure 2e-2h) with asymmetric/symmetric patterns during solstice/equinox seasons, and a globally warmer northern winter (Figure 2g). They seem to appear larger vertical temperature gradients compared to the observations, but two factors, sampling and vertical convolution resulting from the observation configuration, need to be considered for proper model-observation comparisons. The model results should be firstly sampled at the same locations and times with the observations to avoid possible sampling biases, which are however negligible for EMIRS observations due to its near-uniform sampling (Fan et al., 2022b). The other factor is the vertical smoothing of temperature profiles resulting from the near-nadir viewing geometry. The retrieved temperatures at given pressure levels are weighted averages of their respective vertically neighboring levels represented by averaging kernels (black solid lines in Figure 1k-1n). These kernels are computed from the retrieval algorithm (Smith et al., 2022), which need to be applied to the model results before doing comparisons with observations. With these two factors taken into consideration, the results (Figure 2i-2l) show much better agreements but still with systematic differences (Figure 2m-2p). The model generally overestimates the zonal and diurnal mean temperature at low- to mid-latitudes between ±60°, and underestimates it near poles. Such a difference was previously identified in Fan et al., (2022b) during the aphelion season, while here it shows consistency among all seasons (Figure 2m-2p). The overestimates usually maximize at ~100 Pa in the equatorial region, with values of ~6 K in clear seasons (Figure 2m, 2p) and >10 K during the dusty seasons (Figure 2n, 2o). This difference was previously reported in Forget et al., (2022), which suggests a possibly missing process since improvements have been made in simulating the vertical dust profiles (Madeleine et al., 2011; Colaïtis et al., 2013). Near the northern winter solstice (Figure 2o), the temperature is overestimated by >10 K in the downwelling branch of the Hardley circulation near 45°N at ~30 Pa, and underestimated by ~8 K at ~300 Pa. It suggests improvements are needed in the modelled general circulation, which is related to the dust distribution and possibly gravity waves (Gilli et al., 2020; Liu et al., 2023; 2025). Although the numbers of observations in the high-latitude regions are much smaller than those at low latitudes, the model shows consistent underestimates of temperatures in the polar regions of a few Kelvins. Such a difference is more significant during the aphelion season at the winter pole with a value >10 K (Figure 2m), which are likely related to the water and the $CO_2$ cycles, previously identified and suggested by Lefèvre et al., (2008) and Navarro et al., (2014). Multiple processes in the Mars PCM, especially the microphysical processes related to water and $CO_2$ clouds, require further optimizations.

4.3 Diurnal temperature variation
Zonal mean daily temperature anomalies (Figure 3a-3d, 4) are obtained by subtracting the zonal and diurnal means (Figure 2a-2d) from the binned observations. It is the first time that such diurnal temperature variations are observed on a planetary scale for an entire Martian year without significant gaps in local time or influences from seasonal variations. The anomalies in the equatorial region show diurnal cycles in all seasons with one temperature maximum and minimum on each sol (Figure 3a, 3b, 3d), except for the winter solstice when a semi-diurnal cycle appears stronger (Figure 3c). The temperature extrema move downward linearly in the log-pressure coordinate with increasing local time (Figure 3a, 3b, 3d), which is an indication of the downward phase propagation of the dominant diurnal tide, the DW1 mode propagating westward with a period of 1 sol and a zonal wavenumber 1 (Section 3.2). As the phase velocity of diurnal tide has opposite sign to its group velocity, suggested by the tidal theory (Lindzen & Chapman, 1969; Forbes, 1995), the downward phase propagation corresponds to an upward energy transport, which is consistent with the fact that the excitation sources of the tide are near the surface. The range of diurnal temperature variation in the equatorial region is ~4-8 K at 500-5 Pa near summer solstice (Figure 3a), the same as that reported in Fan et al., (2022b), and it becomes ~8 K across all pressure levels during equinox seasons (Figure 3b, 3d). The diurnal cycle becomes stronger when the atmosphere is dustier (Figure 3b), but the pattern breaks near the winter solstice (Figure 3c). In this season during $L_S = 265° - 270°$, the equatorial daily temperature anomaly shows two vertically extended maxima and minima at all pressure levels, while the range of the diurnal variation is <6 K. Such a pattern suggests vertically extended heating sources, mostly dust, during this season, associated with a strong semi-diurnal tide which usually has longer vertical wavelength according to the tidal theory (Lindzen & Chapman, 1969; Forbes, 1995). Outside the equatorial region, the daily temperature anomalies in other latitude bins (not shown) show similar patterns to those in Fan et al., (2022b), which also have larger variations when it is dustier. The temperature anomaly near the winter solstice (Figure 4), however, is not this case. It shows dominant semi-diurnal variations across all latitudes with upward propagations of temperature maxima and minima in pressure regions <100 Pa (Figure 4c, 4d, 4f-4i), especially in the northern hemisphere (Figure 4f-4i). This suggests downward group velocities of the tides, and therefore excitation sources above this region, which are likely high-altitude water ice clouds elevated by regional dust storms (Figure 1b; Wu et al., 2021; Guerlet et al., 2023). The wave decomposition results provide further evidence (see Section 4.4).

Numerical simulations of daily temperature anomalies (Figure 3e-3h) have similar behaviors. They show diurnal patterns in most of the seasons except for the perihelion winter solstice when a semi-diurnal feature dominates (Figure 3g). The ranges of temperature variations are typically 1.5-2 times greater than those in observations (Figure 3a-3d), which becomes much smaller when the sampling strategy and the vertical convolution are considered (Figure 3i-3l). The modelled semi-diurnal feature appearing at ~50 Pa near the summer solstice (Figure 3e) is smoothed (Figure 3i), and the semi-diurnal pattern during winter solstice (Figure 3g) becomes more vertically extended (Figure 3k). This suggests that the results derived using near-nadir observations may underestimate the temperature anomalies and therefore the tide amplitudes by a factor of 1.5-2, consistent with the TES findings (Banfield et al., 2003), and it is more significant if the amplitudes have large vertical variations (Figure 3e, 3i).

4.4 Migrating thermal tides
To investigate the contributions of the atmospheric waves on diurnal temperature variations, amplitudes and phases of these waves (Figure 5, 6) are derived by applying the wave mode decomposition (Section 3.2) onto individual temperature observations at each pressure level in each latitude bin. All prescribed wave modes with integer temporal and zonal frequencies less or equal to 3 are isolated. We show and discuss the migrating thermal tides here, the ones propagating Sun-synchronously, as they typically have the largest amplitudes and the greatest contributions to the diurnal temperature variations in the lower atmosphere <50 km (Banfield et al., 2003; Lee et al., 2009; Fan et al., 2022a; 2022b). Investigations on the non-migrating tides and stationary planetary waves are the subject of future work.

The results of the diurnal tide, DW1, in the equatorial region (Figure 5a-5c, 6b) agree with those inferred from diurnal temperature variations (see Section 4.3). This wave mode is the dominant contributor with an amplitude of 2-4 K near the spring equinox (Figure 5a), which increases to 3-6 K towards the dustier seasons (Figure 5b, 5c). The phase of DW1 shows a downward linear propagation in the log-pressure coordinate, corresponding to a vertical wavelength of ~40 km if assuming a constant scale height of 10 km, which slightly increases towards low pressure levels <20 Pa. Such a vertical wavelength is longer than the ~30 km predicted by the classical tidal theory with a zero-wind assumption (Forbes, 1995; 2020). This is mainly due to the zonal mean zonal wind, which introduces Doppler shift that eastward winds lead to longer vertical wavelengths

(Forbes, 2020). Besides, the zonal mean vorticity, resulting from the shears of these winds, can also modulate the vertical wavelength through wave dispersion relations (Takahashi et al., 2006). The structures of DW1 are similar among these three seasons, and the amplitude is ~1.5 times different, which suggests similarly distributed but stronger forcing at equinoxes than aphelion. In contrast, the structure near the perihelion winter solstice (Figure 6b) is completely different. Except for the near-surface heating dominated high-pressure (>200 Pa) region where the DW1 amplitude agrees with its equinox value of ~6 K (Figure 5b, 5c), it is significantly small (<2 K) at most pressure levels <100 Pa. This is also likely due to the zonal mean vorticity during this dusty season that inhibits the propagation of DW1 (Wu et al., 2017). Also, the vertical wavelength of DW1 is relatively small that may lead to large wave diffusion (Forbes, 2020), which was also reported in the Earth's mesosphere (McLandress 2002).

The semi-diurnal tide, SW2, shows similar structures in the equatorial region among all four selected seasons (Figure 5d-5f, 6e). They also have linear downward phase propagation but with large vertical wavelengths of ~200 km derived from a linear fit of the phase changes with log-pressure. The amplitude of SW2 follows the seasonal variation of dust content in the Martian atmosphere (Figure 1b). It increases from ~1-1.5 K near summer solstice (Figure 5d) to ~1.5-2.5 K, which exceeds DW1 at pressure levels <200 Pa near winter solstice (Figure 6e) . This wave mode with large vertical wavelength is thought to be excited by vertically extended radiative forcings (Wilson & Hamilton, 1996; Kleinböhl et al., 2013), which is dust in most cases except for water ice clouds near aphelion (Wilson et al., 2014; Haberle et al., 2020). The ter-diurnal tide, TW3, is well-constrained in the wave mode decomposition (Figure 5g-5i, 6h) as a result of the good coverage of EMIRS. Such a feasibility was assessed during the aphelion season by Fan et al., (2022b). This mode has vertically extended features without clear vertical propagations except for the aphelion season (Figure 5g). Its vertical wavelength is seemingly larger than that of SW2. Such a behavior suggests that TW3 may also be excited or highly influenced by the atmospheric dust content, and it seems even more sensitive than SW2. The amplitude of TW3 increases from ~0.25-0.5 K in summer (Figure 5g) to ~0.75-1.25 K in winter (Figure 6h).

The wave structures outside the equatorial region during the perihelion season (Figure 6) show features distinct from other times of year. While DW1 has small amplitudes of ~1 K and ~2 K at 30°N and 30°S, respectively (Figure 6a, 6c), it shows upward phase propagation at <100 Pa. Given the opposite directions between phase and group velocities (Forbes, 1995), this upward phase propagation indicates downward group velocity, corresponding to excitation sources above this region at smaller pressure levels. They are probably high-altitude water ice clouds (Wu et al., 2021; Guerlet et al., 2023) elevated by a regional dust storm in the southern hemisphere (Figure 1b). During such a dusty season, SW2 has large amplitudes of ~2 K and ~1.5 K at 30°N and 30°S, respectively (Figure 6d, 6f), and also upward phase propagating features at <100 Pa. The amplitude of SW2 is larger than that of DW1 at 30°N (Figure 6c, 6f), but smaller at the southern latitudes close to regional dust storms (Figure 6a, 6d), which together with the upward phase propagation is likely due to the heating of elevated clouds (Wu et al., 2021). The underlying mechanisms, as well as the vertical distribution of airborne dust, require further investigation. The TW3 at subtropics during this perihelion season (Figure 6g, 6i) has similar vertically extended structures but larger amplitudes (~0.75 K). Such a feature follows the seasonal variation of atmospheric dust content, and agrees with the patterns near the equator.

The numerical simulations show generally good agreement with the observations during non-perihelion seasons in terms of the wave vertical structures, but with differences in their amplitudes and phases (Figure 5, 7). The simulated DW1 (Figure 7a-7c) has a similar linear downward phase propagation feature with a similar vertical wavelength, but its amplitude is ~1 K smaller than the observation at all pressure levels. The phase difference between the model and observation is less than 0.5 hr in local time during all non-perihelion seasons, which is much smaller than the 1.5-3 hr delay shown in Fan et al., (2022a; 2022b) using the old model version (Mars PCM version 5). The smaller amplitude with better phase timing of the modelled DW1 suggests better vertically distributed but underestimated excitation sources in the upgraded PCM, which are mainly airborne dust and water ice clouds. It is the same case for the simulated SW2 (Figure 7d-7f) that good agreements exist in the wave propagation pattern and phase timing compared to observations (Figure 5d-5f), but the amplitude is slightly smaller (~0.5 K). The previously underestimated vertical wavelength of SW2 (Fan et al., 2022b) is also addressed in the upgraded model. Similarly, the modelled TW3 (Figure 7g-7i) has the same phase structures with the observed ones (Figure 5g-5i), but their amplitudes are smaller by a factor of ~2.

In contrast to the non-perihelion seasons, the PCM results near the winter solstice (Figure 8) show significant differences when compared with observations (Figure 6). The DW1 (Figure 8a-8c) have much smaller amplitudes and completely different phases, especially at pressure levels <100 Pa. The simulated DW1 is too weak to be well constrained in the wave mode decomposition. The modelled SW2 (Figure 8d-8f) shows similar vertically extended structures to observations (Figure 6d-6f) but with larger amplitudes, which is significant in the southern hemisphere (Figure 8d). The results of simulated TW3 (Figure 8g-8i), however, show surprisingly good agreement with observations (Figure 6g-6i) in both amplitudes and phases, with the latter denoted by the local times of the vertically extended temperature extrema. In terms of wave propagating directions, only DW1 at southern subtropics shows an upward phase propagating feature (Figure 8a), agreeing with that in the observation (Figure 6a). This suggests elevated forcing within the regional dust storm area. However, such an upward phase propagating feature is not seen in the north (Figure 8c) or SW2 anywhere (Figure 8d, 8f). Improvements in dust and water cycle modelling, especially during the dusty perihelion season, may help to address this discrepancy.

4.5 Latitudinal distribution and seasonal variation of tides
By repeating the wave mode decomposition to each pressure level within each latitude bin, latitudinal and vertical distributions of the tide amplitudes and phases are derived (Figure 9). The DW1 mode shows a mostly symmetric north-south distribution close to the equator during the aphelion season (Figure 9a), which consists of two kinds of patterns in the equatorial region and mid-latitudes, respectively. In the equatorial region between ±30°, the DW1 amplitude peaks near equator with a downward phase propagation (black-blue-white-red in color when moving downward), while those at mid-latitudes maximize poleward of 40° and do not have clear phase changes across different pressure levels. This distribution agrees well with the tidal theory (Lindzen & Chapman, 1969; Forbes, 1995; Wu et al., 2017) that the Hough functions of propagating modes usually have maximal amplitude at or near the equator, and those of the trapped modes, which do not propagate, usually peak around 60°. The distributions of DW1 during the two equinox seasons (Figure 9b, 9d) show similar latitudinally symmetric features to that during the aphelion season, except for a strong (>6 K) trapped mode in the southern hemisphere (40°-60°S), which is likely due to higher atmospheric dust content (Figure 1b). The DW1 is weak at most locations and pressure levels near winter solstice (Figure 9c) due to strong zonal winds and vertically extended dust (see Section 4.4), meanwhile the trapped mode is strong at large pressure levels (>100 Pa) in the south (0°-60°S). Some upward features (black-blue-white-red in color when moving upward) are observed outside the equatorial region, agreeing with those discussed in Section 4.4.

The amplitude and phase distributions of SW2 (Figure 9e-9h) are not as clearly coherent as DW1. Only during summer and autumn can an amplitude maximum be identified in the equatorial region across all pressure levels along with downward phase propagation (Figure 9e, 9f). The respective asymmetric and symmetric patterns during these two seasons are likely due to the latitudinal distribution of dust and the large-scale circulation. In contrast, the results during the other two seasons (Figure 9g, 9h) do not show such clear latitudinal distributions of SW2, which may likely be due to complex atmospheric conditions during the dustiest and the dust-declining seasons. The amplitude of SW2 is generally large when the dust content is high, but it does not show a regular pattern. The distributions of TW3 (Figure 9i-9l) are more loosely constrained. Similar to SW2, TW3 is generally strong during dusty seasons. A clear amplitude maximum is only seen during the northern winter (Figure 9k). Detailed excitation sources of TW3 and their relative importances are yet to be discovered. Non-linear interactions between waves, especially during the dusty season when a number of them are strong, is a possible candidate.

The latitudinal-vertical structures of simulated migrating thermal tides (Figure 10), where the sampling and vertical convolution are considered, show qualitative agreements with the observations (Figure 9). The modelled DW1 captures the seasonal variations well but with noticeable differences in amplitude and phase. It has symmetric patterns during non-perihelion seasons (Figure 10a, 10b, 10d), which feature the equatorial downward-phase-propagating modes and the mid-latitude constant-phase trapped modes. The strong southern trapped mode at 40°-60°S in equinoctial seasons is also seen (Figure 10b, 10d). Around the winter solstice, the simulated DW1 amplitude (Figure 10c) shows a similar pattern as observations but with >1 K larger values (Figure 9c), which are instead ~1 K smaller during the other three seasons. The phase propagation of the simulated DW1 in the equatorial region agrees with that observed during seasons except the winter solstice when it is not so credible given the small <1 K amplitude. The overall seasonal trend of SW2 is captured by the model with larger amplitudes when the atmosphere is dustier (Figure 10e-10h), but its amplitude distribution shows large disagreements. The simulated SW pattern consistently shows equatorial downward-phase-propagating Hough modes during all four selected seasons, which is not the case observed. Same as DW1, the amplitude of the simulated SW2 is larger than the observed value near the winter solstice (Figure 10g), but it is the other way in the other three seasons (Figure 10e, 10f, 10h). Such discrepancies indicate that special treatment may be needed towards improving the modelled dust vertical distribution and its temporal variation. The model-observation agreement of TW3 is partial and limited. The model results only show the importance of TW3 near perihelion (Figure 10k), which is mostly <0.25 K during the other three seasons. In contrast, TW3 is observed to be double the amplitude during summer and autumn (Figure 9i, 9j, 10i, 10j), and ~0.25 K stronger in winter (Figure 9k, 10k). Further investigations on the excitation mechanisms of TW3 are expected.

The latitudinal distribution of DW1 amplitude shows an annual cycle (Figure 11a-11c). Its equatorial maximum increases from aphelion to dusty seasons but has a minimum near the dustiest perihelion, which agrees with the discussion above (see Section 4.4). The amplitude distribution varies latitudinally at each pressure level with the center of the equatorial propagating mode moving from ~5°S in summer to ~5°N in winter at 100 Pa (Figure 11a), while ~15°S to ~15°N at 10 Pa (Figure 11c). It crosses the equator twice but reaches each mid-latitude once during this MY, leading to a semi-annual cycle in the equatorial region and an annual cycle at mid-latitudes, which was previously reported in the MCS data (Wu et al., 2017). The two mid-latitude temperature maxima of the trapped modes have similar seasonal variations, but their latitudes are not so well constrained because of small observation numbers at high latitudes (Figure 1g-1j). Such latitudinal distributions and their seasonal variations are well simulated by the Mars PCM (Figure 11d-11f), which shows good agreement when including the observation sampling and vertical convolution (Figure 11g-11i) despite the smaller simulated tide amplitudes. It is worth to be noted that the results at 10 Pa are meaningful, as the sum of the kernel elements is usually >0.8 at this pressure level and it keeps the seasonal variation (Figure 11f, 11i). The seasonal meridional movement of the tide amplitude distribution can be well explained by the zonal mean total vorticity, the sum of the Coriolis parameter and the horizontal shear of zonal mean zonal wind. This mechanism has been successfully used to explain the seasonal variations of tides in the Earth's mesosphere (McLandress, 2002) where the zonal mean vorticities can lead to varying effective planetary rotations within the context of atmospheric dynamics. As the atmospheric waves are oscillations on top of mean flows, they "feel" differential planetary rotations in this "dynamical" coordinate introduced by the zonal mean zonal wind. The "dynamical equator" (white dashed lines in Figure 11) migrates with season, influenced by the advection of momentum due to the Hadley circulation. The equatorial propagating mode of DW1 has its amplitude maximum aligning well with this "dynamical equator", which changes with pressure levels and seasons. Larger latitudinal variations are seen at lower pressure levels where zonal winds are stronger (Figure 11c). This phenomenon was not predicted in the classic tidal theory (Lindzen & Chapman, 1969; Forbes, 1995) where zero zonal mean flows are assumed. It suggests the necessity of including the zonal wind into the analysis of tides in the Martian atmosphere, potentially following McLandress, (2002).

The SW2 has different seasonal patterns than DW1 with worse model-observation agreements (Figure 12). The annual cycle of the observed SW2 amplitude generally follows that of the atmospheric dust content (Figure 1b), but with substantial differences. The SW2 amplitude peaks during three regional dust storm periods around $L_S$ = 155°, 230°, and 315°, but it also remains large between the latter two storms during $L_S$ = 270° – 300°. Such a feature is not seen in the original simulation (Figure 12d-12f) or when the observation sampling and convolution are considered (Figure 12g-12i). The simulated SW2 amplitude has similar values as observations during dust storms, but it decays much faster afterwards to smaller values during non-storm seasons. This may be due to faster dust removing processes and thus weaker radiative forcings in the PCM. In contrast to the observations, the annual cycle of the simulated SW2 (Figure 12g-12i) aligns well with the modelled local dust mixing ratio (white contours in Figure 12), apart from an amplitude peak due to water ice clouds (not shown) during $L_S$ = 90° – 120°. The seasonal patterns of SW2 are distinct at different pressure levels. The equatorial amplitude maximum moves from ~20°N to ~20°S from aphelion to perihelion at 100 Pa (Figure 12a), while it is the other way from south to north at 10 Pa (Figure 12c). This may be due to different airborne dust mixing ratios at these pressure levels influenced by the atmospheric circulation, which is well simulated by the PCM (Figure 12g-12i).

The TW3 is observed to have large seasonal variations which are poorly captured by the numerical simulation (Figure 13). The latitudinal distribution of its amplitude shows an equatorial maximum, which appears to follow the atmospheric dust content, and seems more sensitive than that of SW2. The amplitude of TW3 increases from <0.5 K in the aphelion season to >1.5 K during regional dust storms (Figure 13a-13c), and keeps at a large value between the A- and C-storms during $L_S = 270° – 300°$, similar to that of SW2 (Figure 12a-12c). However, TW3 does not show a large amplitude in the aphelion season during $L_S = 90° – 120°$ when the water ice cloud belt takes place, which suggests it is likely controlled by dust instead of clouds. Also, the seasonal change of the TW3 equatorial maximum moves across a large latitude range at 100 Pa, from ~20°N during aphelion to ~20°S during perihelion (Figure 13a), but there is not such a clear change at 10 Pa (Figure 13c). This reinforces the scenario that TW3 is likely excited by processes in the lower atmosphere. The Mars PCM captured the general trend of TW3 with increasing amplitude while the atmosphere became dustier, especially during the A- and C-storms at $L_S = 230°$, and 315°, respectively (Figure 12d-12f). However, the sampled and vertically convolved TW3 has smaller amplitude than the observation (Figure 12g-12i), especially at large pressure levels (Figure 12g). Such a difference indicates missing excitation processes of TW3 in the lowermost layers in the model. Similar to SW2, the modelled TW3 decays too fast after dust storms, pointing to improvements in the dust removing process in the Mars PCM.

5.  **Discussion**

Diurnal temperature variations and migrating thermal tides in the Martian atmosphere are analyzed using EMM/EMIRS observations during its primary mission. Such diurnal variations are observed for an entire Martian year on a planetary scale without significant gaps in local time or influences from seasonal variations. The observations are binned 5° in $L_S$ in these analyses, which is the minimal required time for the instruments to complete a full coverage of both location and local time. This $L_S$ range is sufficient to derive the diurnal temperature anomalies. Results during the calm aphelion season are almost identical with those in Fan et al., (2022b) where the data are binned 30° in $L_S = 60° – 90°$. This range is also good for the dusty seasons as long as the onsets of dust storms are not included, which is indicated by the smooth changes of the daily temperature anomaly pattern from one $L_S$ bin to the next. However, the 5° $L_S$ is not long enough for deriving stable wave decomposition with good statistics. The total observation number of ~2000 in the equatorial region when using 5° $L_S$ bin sizes is barely sufficient for the linear regression with 49 degrees of freedom, but the performance decreases sharply towards mid-latitudes with only hundreds of observations within each 10° latitude bin (Figure 1g-1j). An example of the decomposition result using only the data in each 5° $L_S$ bin shows poor statistical stability (Figure 14). The derived amplitudes of the migrating tides present rapid jumps from one $L_S$ bin to another, which indicates overfitting. Also, the latitude range where the tides can be well constrained is much narrower, typically <45° for SW2 and ~30° for TW3. Therefore, a 15° $L_S$ window length is necessary and therefore chosen for the wave mode decomposition, which may include some real seasonal changes during the perihelion season or during dust storms. Given the fact that the time period required by EMIRS to complete a full coverage of location and local time is already short, significant improvements may not be possible if only using observations from a single instrument. Constellations with carefully designed synergetic observation strategies may help.

As a result of the low EMM orbit inclination, more observations are obtained in the equatorial region and subtropics, which results in successful analyses of diurnal temperature variations and atmospheric waves. However, the tidal theory suggests that many modes of migrating tides should have large amplitudes at mid- to high-latitudes (Lindzen & Chapman, 1969; Forbes, 1995), which are outside the region where the waves can be well constrained using EMIRS observations alone. Therefore, the Hough modes of each migrating tide are not further decomposed.

The full and near-uniform local time coverage of EMIRS observations largely addressed the aliasing issue, which was significant in previous investigations using MGS/TES (Banfield et al., 2003) or MRO/MCS (Lee et al., 2009) data. Although this issue was partially relieved by including MCS cross-track measurements, the derived uncertainties of some wave modes, e.g., SW2, were still large (Kleinböhl et al., 2013; Wu et al., 2015; 2017; 2021). The EMIRS results show good agreement with those derived from MCS limb soundings if the vertical convolution is considered, which usually decreases the observed wave amplitudes by a factor of 1.5-2. Wu et al., (2015; 2017) reported an 8 K maximum of DW1 in the equatorial region at tens of pascals using MCS observations through two MYs, which is within the 1.5-2 times of the EMIRS results (Figure 7). In addition, the region with a strong trapped DW1 >8 K at >10 Pa and 40°-60°S is also seen in MCS observations near autumn equinox and during dust storms (Wu et al., 2021), which is >6 K in this work (Figure 9b-9d). Kleinböhl et al., (2013) and Wu et al., (2015) presented a 4-8 K SW2, although with large uncertainties due to unevenly distributed observations, in the equatorial region at tens of pascals through the entire MY using MCS data binned between ±20° and ±10° in latitude, respectively. This amplitude is approximately twice that in this work (Figure 7, 8), agreeing with the vertical smoothing of averaging kernels.

The tide propagating features presented in this work also agree with previous observations. The vertical wavelengths of DW1 and SW2, ~40 km and ~200 km, respectively, in the lower ~40 km agree with those derived using TIRVIM data (Fan et al., 2022a). Forbes et al., (2020) reported a ~50 km DW1 wavelength observed by MCS at 20-70 km, which coincides with the <20 Pa region in this work where the DW1 wavelength starts to increase while propagating upward. The longer vertical wavelength of SW2 was also seen in the MCS observations (Kleinböhl et al., 2013), which, however, was not well constrained at pressure levels >50 Pa due to the aforementioned uneven sampling. During dust storms, the upward phase propagation of tides and the small DW1 amplitude observed in EMIRS data (Figure 6) agree with those seen by MCS (Wu et al., 2021) and TIRVIM (Guerlet et al., et al., 2023), providing further evidence of possible high-altitude excitation sources, likely elevated water ice clouds. Investigations combining these results with EMIRS dust and water ice retrievals may have great significance.

In terms of seasonal variations, Wu et al., (2017) presented a semi-annual cycle of DW1 at the equator and an annual cycle at mid-latitudes (~45°). Results from this work suggest that both of these two cycles are from the same annual cycle of DW1. Its equatorial maximum crosses the equator twice and reaches mid-latitudes once per MY (Figure 10), which is well explained by the total zonal mean vorticities (see Section 4.5). Using MCS data from multiple MYs, Forbes et al., (2020) analyzed the climatological behaviors of the DW1 and the SW2 propagating modes at 76 km altitude, where these tides are expected to have the largest amplitudes. Their results show a similar seasonal behavior of DW1 in the middle atmosphere. The DW1 amplitude generally increases when the atmosphere is dustier and presents an aphelion season maximum due to water ice clouds, same as those in this work (Figure 11a-11c). The weak amplitude during the perihelion was also observed, which decreases near $L_S = 180°$ and returns at $L_S = 300°$. Nevertheless, the latitudinal distribution shows noticeable differences, which is likely due to distinct wind patterns at different vertical levels. The SW2 amplitude shows substantial differences between the lower and the middle atmospheres. The MCS observations suggest a weak equatorial maximum of SW2 at 76 km altitude near $L_S = 270°$ between the two regional dust storms, which, in contrast, remains large in the lower atmosphere (Figure 12a-12c). Such a difference may be due to the fact that the large SW2 amplitude seen in the EMIRS analysis is mainly composed of trapped Hough modes, which are excited by local airborne dust and do not propagate upward.

The atmospheric tides are also observed in the near-surface pressure field measured by landers and rovers (e.g., Sánchez-Lavega et al., 2022; Leino et al., 2023; 2024). Due to their high temporal sampling frequencies, these observations were used to resolve up to six tidal components (Sánchez-Lavega et al., 2022). These tides in pressure measurements are mainly controlled by near-surface heating and highly influenced by the topography. Therefore, they are ideal for distinguishing the altitudes of excitation sources if compared with atmospheric observations. Due to the local-scale spatial coverage, aliasing of waves with the same temporal frequency but different zonal wavenumbers make it difficult to isolate all tidal components. Nevertheless, the seasonal trends of the diurnal and the semi-diurnal pressure oscillations generally follow the atmospheric dust content (Sánchez-Lavega et al., 2022; Leino et al., 2023), similar to the behaviors of the DW1 and the SW2 migrating tides in the atmosphere (Figure 12, 13). The surface pressure oscillations always show amplitude spikes during regional dust storms, while the DW1 in atmospheric temperature is observed to be small when very dusty. This confirms the scenario that vertically extended airborne dust damps DW1 and prevents its propagation. The ter-diurnal surface pressure oscillations observed by the landers and rovers show larger differences from the model predictions (Leino et al., 2024), similar to the TW3 behaviors shown in the EMIRS data (Figure 14). The oscillation in surface pressure is suggested to follow both the atmospheric dust content and water ice clouds, but the EMIRS TW3 results indicate that water ice clouds may not be important (Figure 13a-13c). Improvements in the numerical simulation of TW3 are necessary for a more detailed interpretation and a better understanding.

Thermal tides are sensitive indicators of their excitation sources and transmission media, which are typically major radiative forcings in the Martian atmosphere, e.g., the near-surface atmosphere, topography, airborne dust, water and $CO_2$ ice. The amplitudes of these tides usually reflect the strengths of their forcings, while the phases indicate corresponding vertical distributions. The good model-observation agreements of the DW1 phase and the SW2 wavelength (Figure 6-9) indicate successful upgrades of physical processes in the Mars PCM (version 6), mainly the vertical distributions of dust and water ice. Disagreements between the model and observations point to areas for in model improvements. The smaller simulated amplitudes of all tides and lower zonal and diurnal mean temperatures (Figure 2m-2p) suggest underestimated forcings in the model. The significant differences of diurnal mean temperature and temperature anomaly during dusty seasons (Figure 3, 6, 8), as well as the direction of the tide phase propagation, are likely due to the different characteristics of the dust cycles. This also suggests thicker or stronger elevated water ice clouds, especially in the subtropics. Moreover, the seasonal variation of the observed SW2 indicates a slower decay pace after regional dust storms than the simulation, which suggests the dust removing processes to be improved.

## 6. Conclusion

This work presents an investigation of diurnal temperature variations and migrating thermal tides in the Martian atmosphere on a planetary scale over an entire Martian year using EMM/EMIRS observations. The observations obtained from the high-altitude orbit can achieve a full geographical and local time coverage within 9-10 sols with near-uniform spatial and temporal sampling. It facilitates detailed investigations of diurnal variations in the Martian atmosphere, and analyses of atmospheric waves without aliasing.

The EMIRS observations show that the Martian atmosphere is warmer with larger diurnal variations when dustier. The daily temperature anomaly shows a diurnal cycle as large as 8 K in most seasons except for the dusty perihelion when the temperature vertical gradient is small and a semi-diurnal pattern dominates.

The diurnal migrating tide, DW1, dominates the diurnal temperature variation around the entire MY except for the northern winter. It becomes stronger with increasing atmospheric dust content, but shows a small amplitude when the atmosphere is dustiest, likely due to strong circulations and vertically extended excitation sources (e.g., airborne dust). The vertical wavelength of DW1, modified by the zonal wind, is ~40 km at tens to hundreds of pascals, slightly larger than that predicted in the classical tidal theory with a zero-wind assumption. The latitudinal distribution of DW1 shows a north-south symmetry with an equatorial maximum of propagating Hough modes and two mid-latitude ones due to trapped modes. The latitudinal distribution of DW1 shows an annual cycle, which is well explained by the total vorticity introduced by the mean zonal wind. It presents the importance of including wind patterns in the analysis of thermal tides.

The semi-diurnal migrating tide, SW2, shows a vertically extended structure with a large wavelength. Its amplitude is usually second to DW1 but surpasses it during northern winter. The seasonal trend of SW2 generally follows the dust content in the Martian atmosphere, besides a peak during the aphelion season due to water ice clouds. A slower seasonal decay of SW2 than the atmospheric dust content is observed between two regional dust storms. The latitudinal distribution of SW2 also shows an annual cycle and is different across pressure levels, which is likely regulated by the distributions of airborne excitation sources as well as large-scale circulations.

The ter-diurnal migrating tide, TW3, is constrained using EMIRS observations as a result of its good spatial and temporal coverage. This tide mode shows a similarly large vertical wavelength as SW2. However, the latitudinal distribution of TW3 is not well determined due to its small amplitude, which only appears important with an amplitude >1 K during the dusty perihelion season. The seasonal variation of TW3 seems sensitive to the atmospheric dust content and also excitation sources in the lowermost atmosphere layers, but not correlated to water ice clouds.

The upgraded Mars PCM (version 6) shows generally good model-observation agreements but with substantial differences. It captured most latitudinal and seasonal trends in the Martian atmosphere while the detailed structures need to be improved. The model consistently overestimates the mean atmospheric temperature up to 6-10 K at <60° latitude, but underestimates it over the poles, indicating improvements needed in simulating the water and $CO_2$ clouds. The diurnal temperature variations are underestimated in the simulation, which leads to the underestimates of all decomposed tides. Specifically, the model usually underestimates DW1 and SW2 by ~1 K and ~0.5 K, respectively, and TW3 by a factor of 2, which suggests updates in the radiative forcings of their excitation sources. The model reproduced DW1 well in terms of its latitudinal structure and seasonal variation, with the annual cycle well explained by the simulated zonal wind. The simulated SW2 has good agreement in the overall seasonal trend, but its latitudinal distribution has large mismatches, which is likely related to the dust distribution. The current model can not simulate TW3 well, asking the need for investigations and interpretations on its excitation mechanisms. These model-observation discrepancies provide valuable information guiding the model improvements in the tide excitation sources and their radiative forcings, mainly dust and ices, as well as temporal behaviors including dust removing processes.

**Acknowledgments**
Funding for development of the EMM mission was provided by the United Arab Emirates (UAE) government, and to co-authors outside of the UAE by the Mohammed bin Rashid Space Centre (MBRSC). This project has received funding from the EU's Horizon Europe research and innovation funding programme under the Marie Skłodowska-Curie grant agreement No. 101064814. S.F. acknowledges funding from the National Natural Science Foundation of China through grant No. 42475133, and the Stable Support Plan Program for the Higher Education Institutions of the Shenzhen Science and Technology Innovation Commission through grant No. 20231115103030002. Computations were partly supported by the Center for Computational Science and Engineering of the Southern University of Science and Technology.

**Open Research**
Data from the Emirates Mars Mission (EMM) are freely and publicly available on the EMM Science Data Center (SDC, http://sdc.emiratesmarsmission.ae). This location is designated as the primary repository for all data products produced by the EMM team and is designated as long-term repository as required by the UAE Space Agency. The data available (http://sdc.emiratesmarsmission.ae/data) include ancillary spacecraft data, instrument telemetry, Level 1 (raw instrument data) to Level 3 (derived science products), quicklook products, and data user guides (https://sdc.emiratesmarsmission.ae/documentation) to assist in the analysis of the data. Following the creation of a free login, all EMM data are searchable via parameters such as product file name, solar longitude, acquisition time, sub-spacecraft latitude & longitude, instrument, data product level, and etc. Emirates Mars Infrared Spectrometer (EMIRS) data and user guides are available at: https://sdc.emiratesmarsmission.ae/data/emirs.

Data products can be browsed within the SDC via a standardized file system structure that follows the convention:
/emm/data/<Instrument>/<DataLevel>/<Mode>/<Year>/<Month>
Data product filenames follow a standard convention: emm_<Instrument>_<DataLevel><StartTimeUTC>_<OrbitNumber>_<Mode>_<Description>_ <KernelLevel>_<Version>.<FileType>

The Mars PCM outputs from MY 36 $L_S = 0°$ to MY 37 $L_S = 60°$ (Fan, 2024) are available on Zenodo with doi: 10.5281/zenodo.14173356. Permission is granted to use these datasets in research and publications with appropriate acknowledgements.

Figures

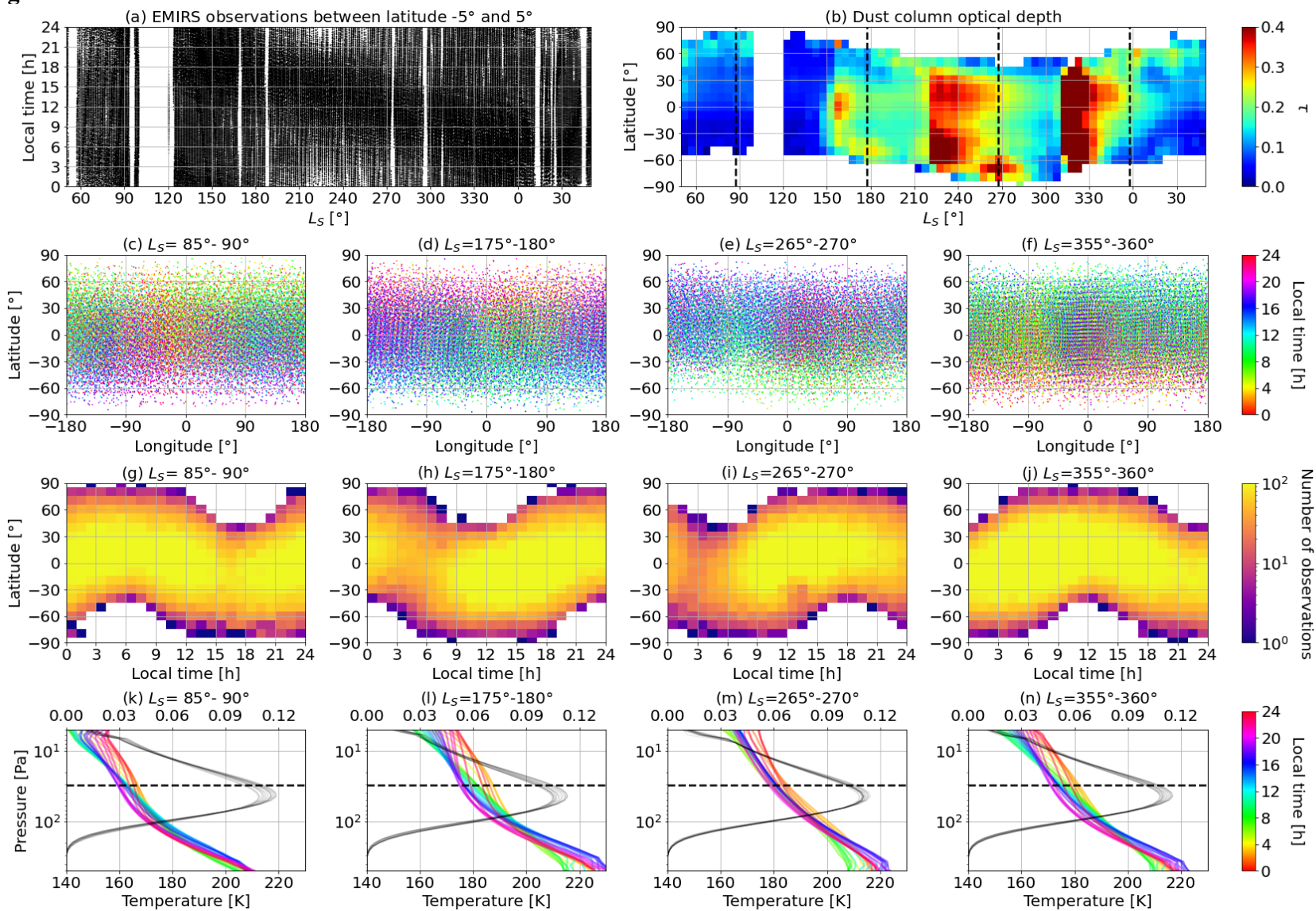

**Figure 1.** (a) Local time sampling of EMIRS observations in the equatorial region between 5° S and 5° N during the EMM primary

mission. (b) Diurnal mean dust column optical depth retrieved from EMIRS observations. The vertical dashed lines denote the four selected seasons used for analysis in this work. (c-f) Location of EMIRS observations during the four selected seasons. Colors of the points denote the observation local time. (g-j) Zonal total numbers of successful EMIRS retrievals in the (latitude, local time) bins during the four selected seasons. (k-n) Zonal mean temperature profiles (color lines and the bottom axis) and typical averaging kernels at 30 Pa (black solid lines and the top axis) in the equatorial bin between 5° S and 5° N during the four selected seasons. Colors of the temperature profiles denote the local time of the temperature profiles, and the horizontal dashed lines denote the 30 Pa pressure level. The pressure level of the retrieved temperatures are averages weighted by corresponding asymmetric averaging kernels, so they do not coincide with kernel maxima pressure levels.

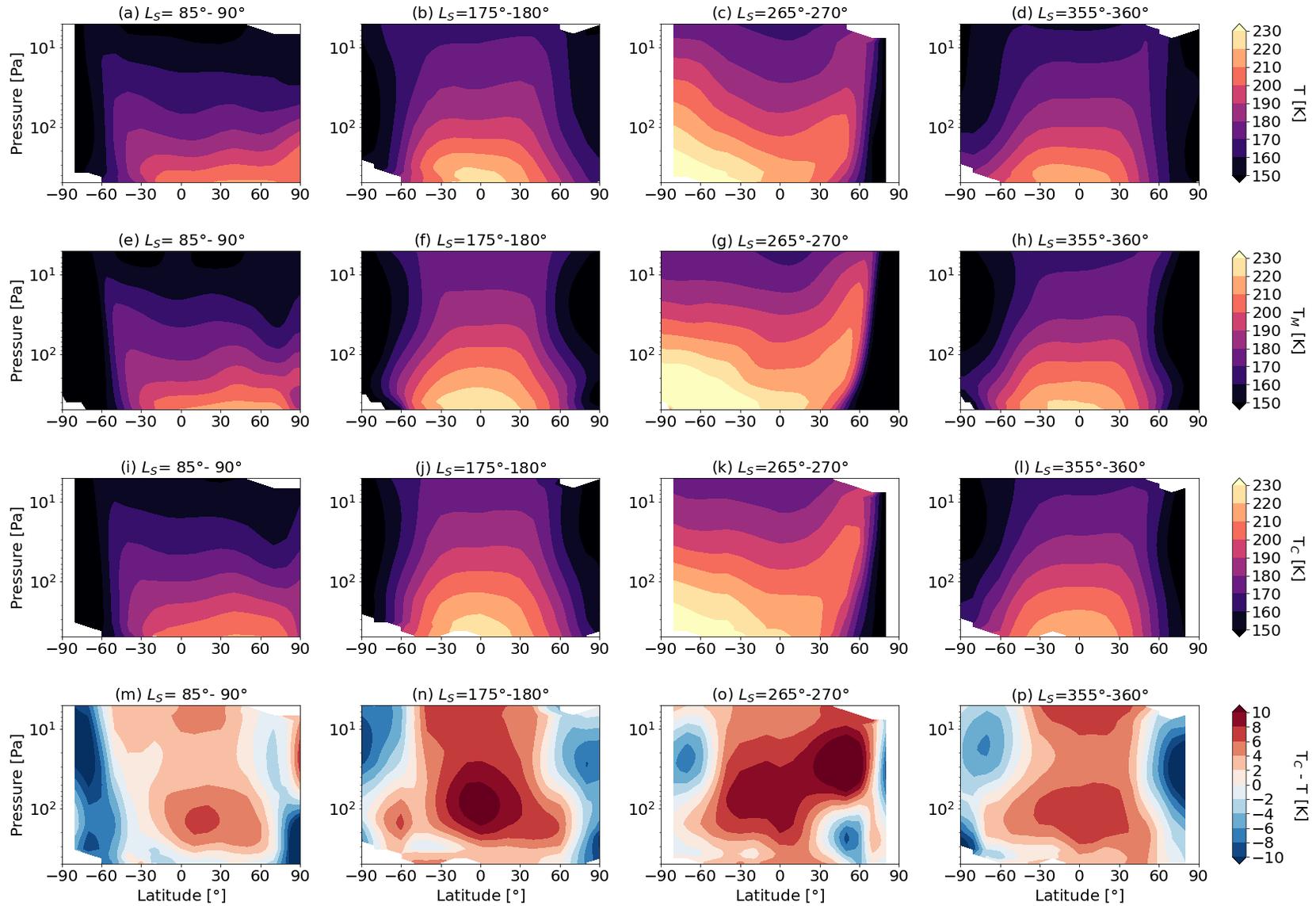

**Figure 2.** (a-d) Zonal and diurnal mean temperatures derived using EMIRS observations (T) within the four $L_S$ bins of the selected seasons at MY 36 $L_S = 85° − 90°, 175° − 180°, 265° − 270°$, and $355° − 360°$, respectively. The contours have intervals of 10K. (e-h)

Same as (a-d), respectively, but for the Mars PCM outputs ($T_M$). (i-l) Same as (a-d), respectively, but for the Mars PCM results sampled at the same locations and times as EMIRS observations, and vertically convolved using kernels derived from the EMIRS retrieval algorithm ($T_C$). (m-p) Model-observation comparisons of the zonal and diurnal mean temperatures, computed using the EMIRS observations (T) in (a-d), and their respective sampled and vertically convolved Mars PCM outputs ($T_C$) in (i-l). The contours have intervals of 2K.

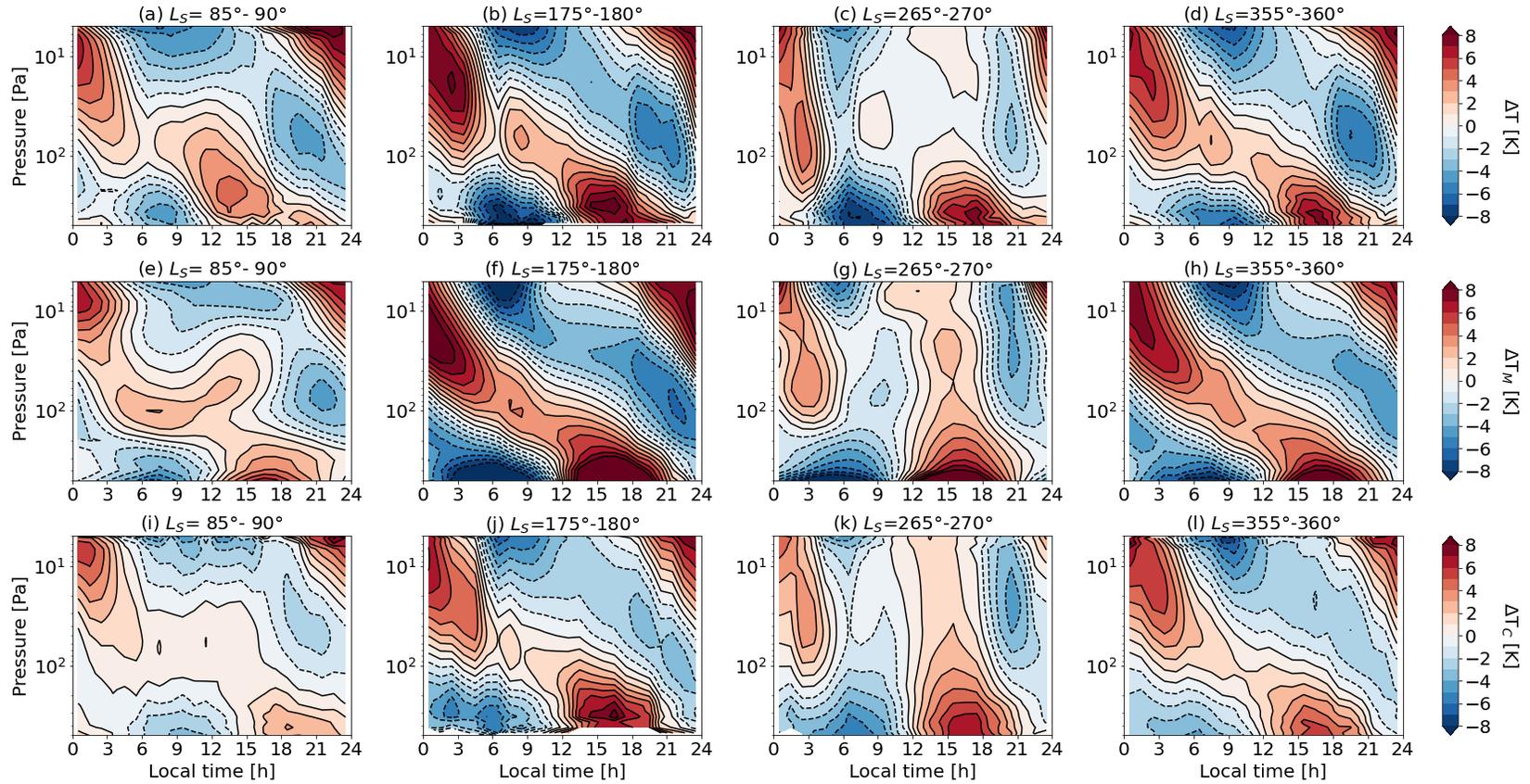

**Figure 3.** (a-d) Zonal mean daily temperature anomalies derived using EMIRS observations (T) in the equatorial bin between 5° S and 5° N within the four $L_S$ bins of the selected seasons at MY 36 $L_S$ = 85° – 90°, 175° – 180°, 265° – 270°, and 355° – 360°, respectively. The contours have intervals of 1K. (e-h) Same as (a-d), respectively, but for the Mars PCM outputs ($T_M$). (i-l) Same as (a-d), respectively, but for the Mars PCM results with sampling and vertical convolution ($T_C$).

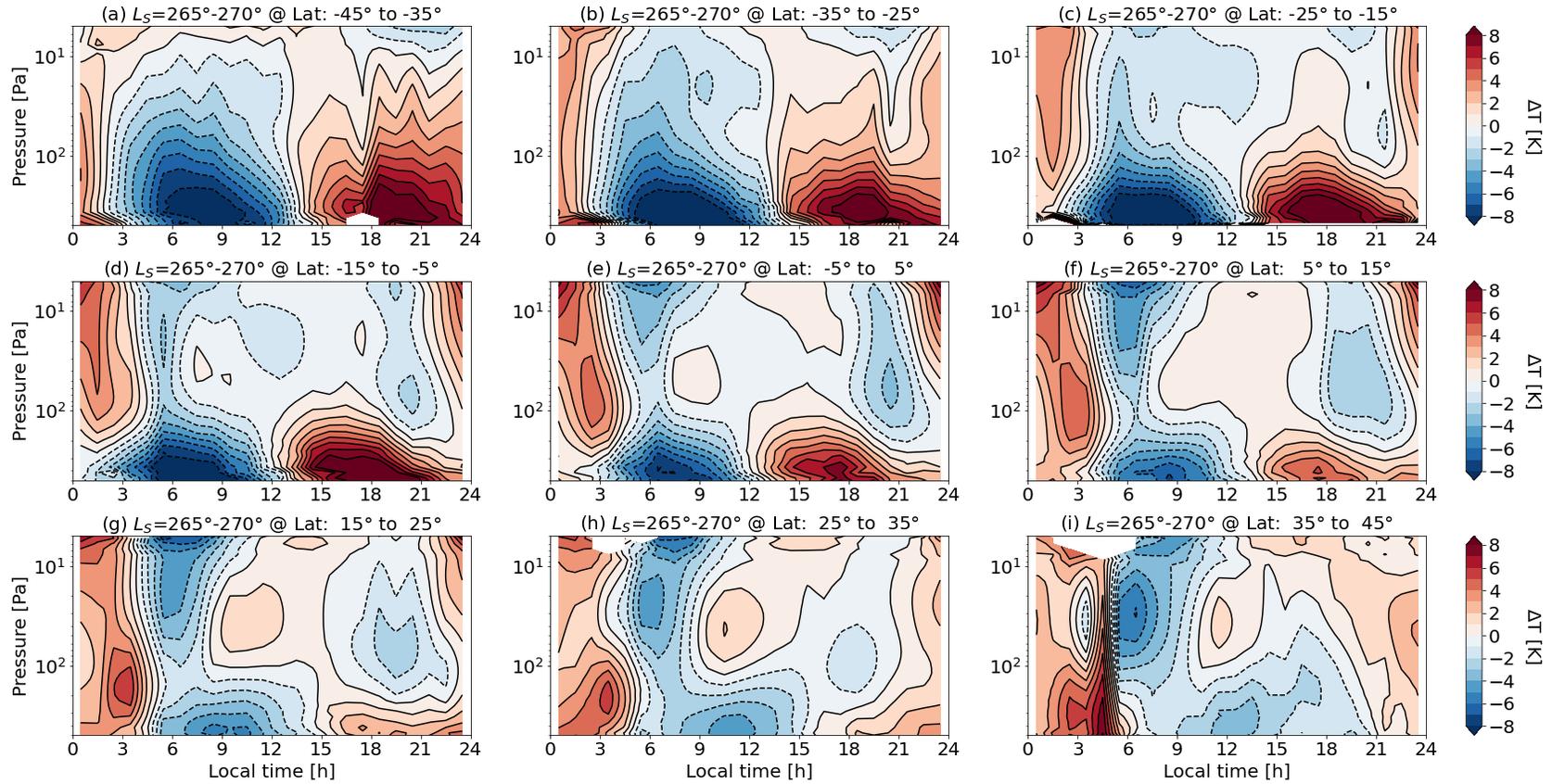

**Figure 4.** Zonal mean daily temperature anomalies derived using EMIRS observations (T) in latitude bins centered at 40°S to 40°N with an interval of 10° within the $L_S$ bin near the northern summer solstice at MY 36 $L_S = 265° - 270°$.

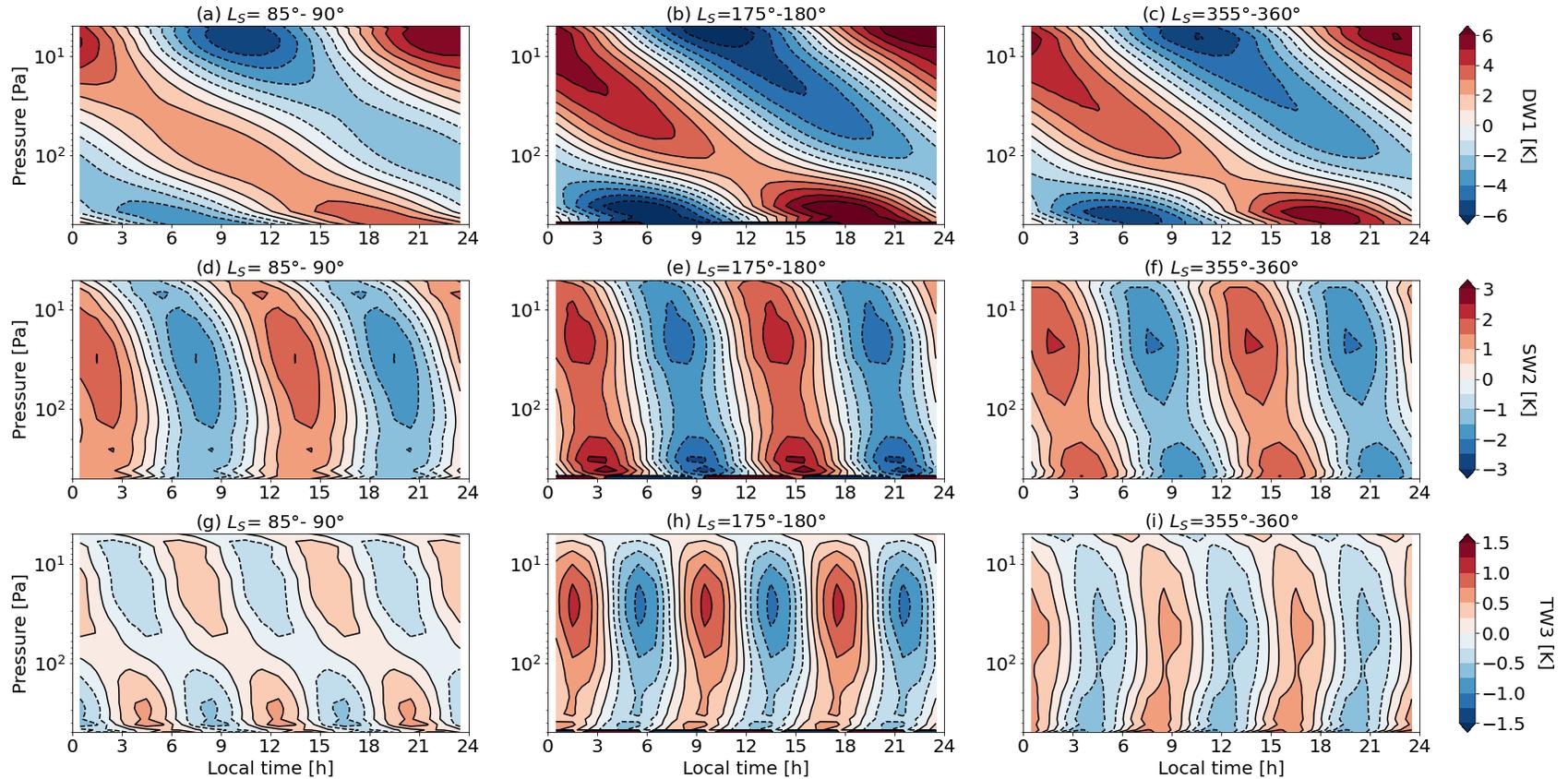

**Figure 5.** (a) The diurnal tide (DW1) derived through the wave mode decomposition using EMIRS observations (T) in the latitude bin between 5° S and 5° N within the $L_S$ bin at MY 36 $L_S = 85° – 90°$ near the northern summer solstice. The contours have an interval of 1 K. (b) Same as (a), but for the $L_S$ bin at MY 36 $L_S = 175° – 180°$ near the autumn equinox. (c) Same as (a), but for the $L_S$ bin at MY 36 $L_S = 355° – 360°$ near the spring equinox. (d-f) Same as (a-c), respectively, but for the semi-diurnal tide (SW2). The contours have an interval of 0.5 K. (g-i) Same as (a-c), respectively, but for the ter-diurnal tide (TW3). The contours have an interval of 0.25 K.

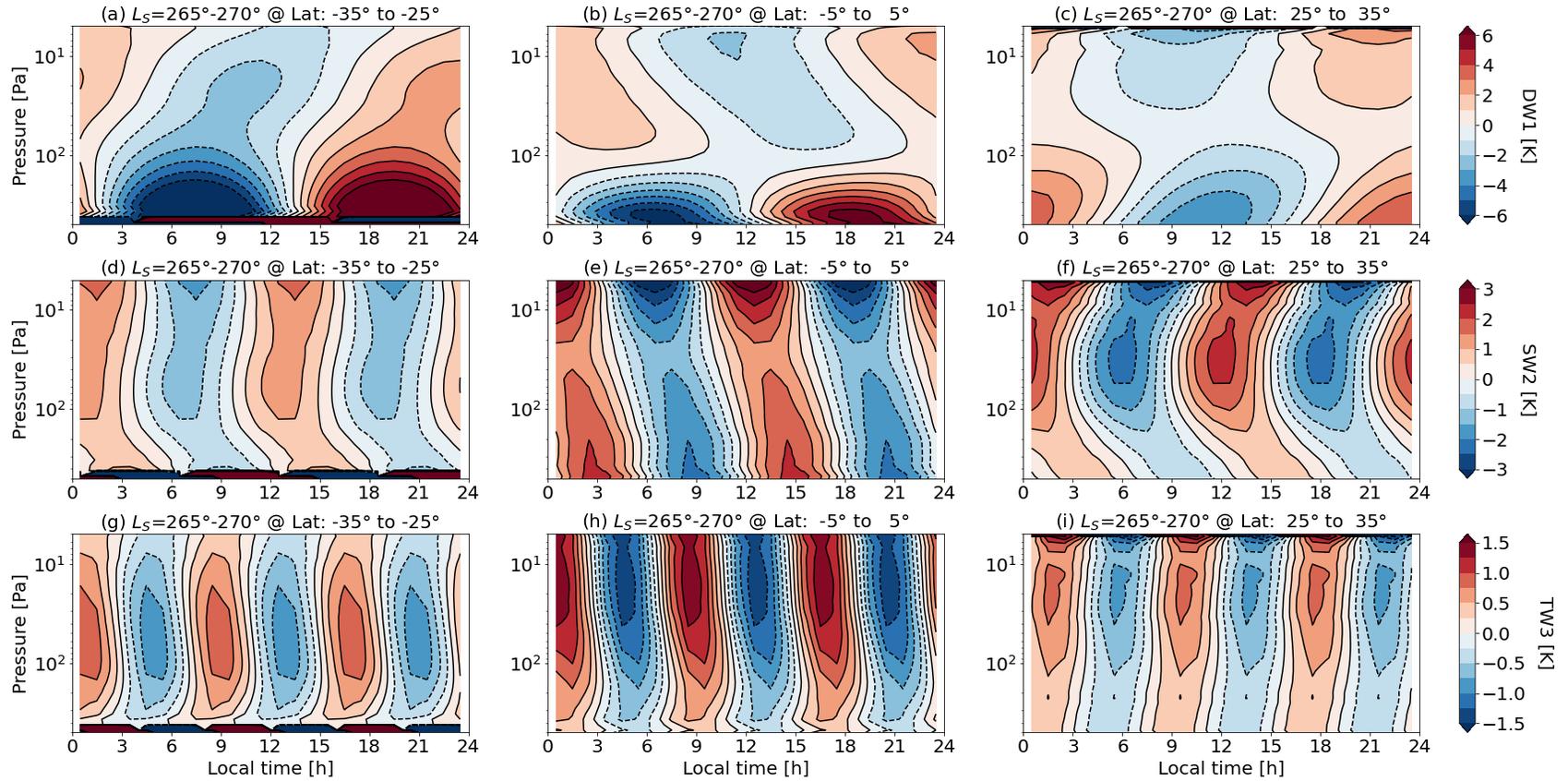

**Figure 6.** (a-c) Same as Figure 5a-5c, but for the latitude bins at 35° S – 25° S, 5° S – 5° N, and 25° N – 35° N, respectively, in the $L_S$ bin at MY 36 $L_S = 265° - 270°$ near the winter solstice. (d-f) Same as (a-c), respectively, but for the semi-diurnal tide (SW2). (g-i) Same as (a-c), respectively, but for the ter-diurnal tide (TW3).

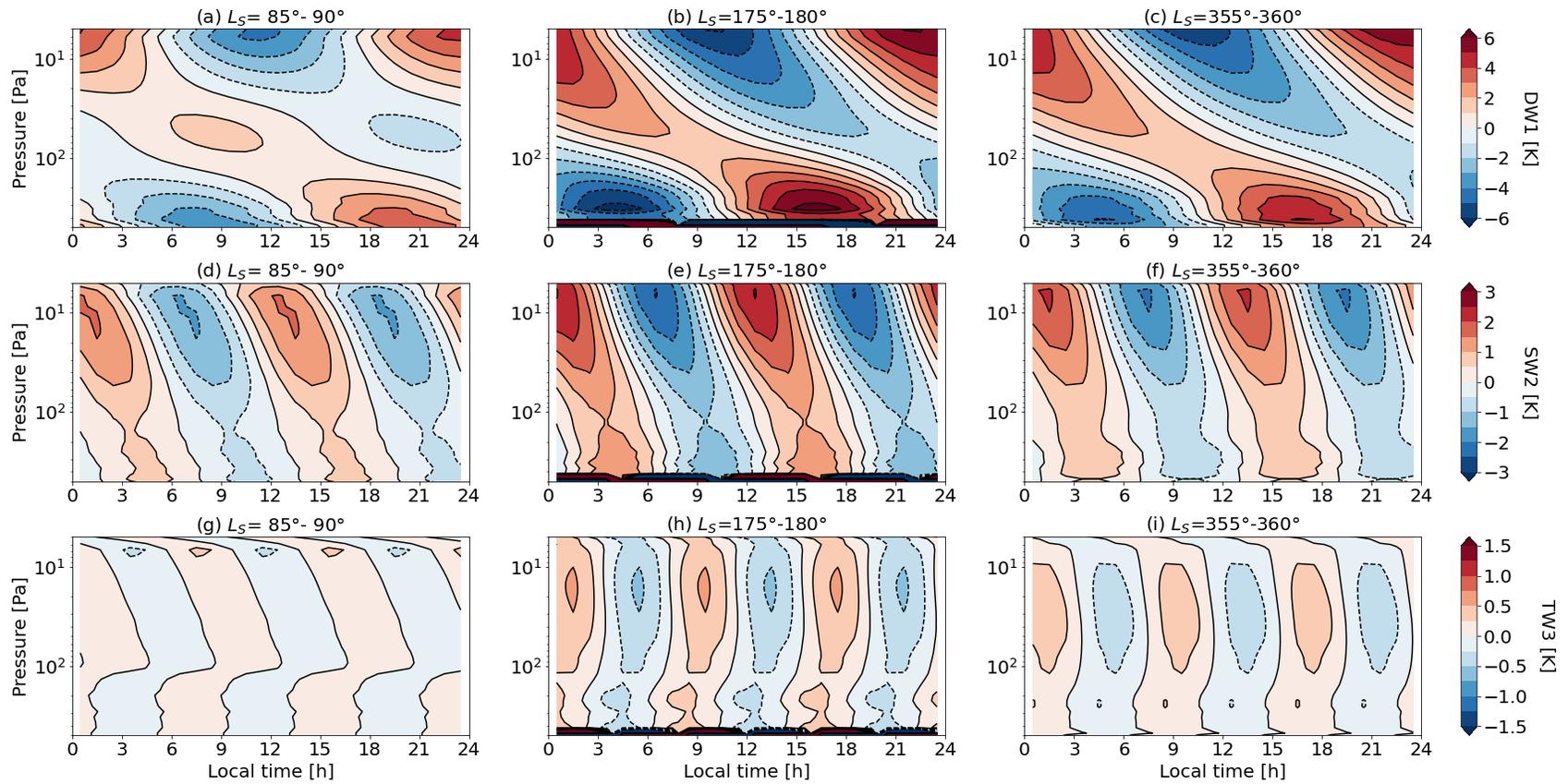

**Figure 7.** Same as Figure 5, but for the decomposition results using sampled and vertically convolved Mars PCM outputs ($T_C$).

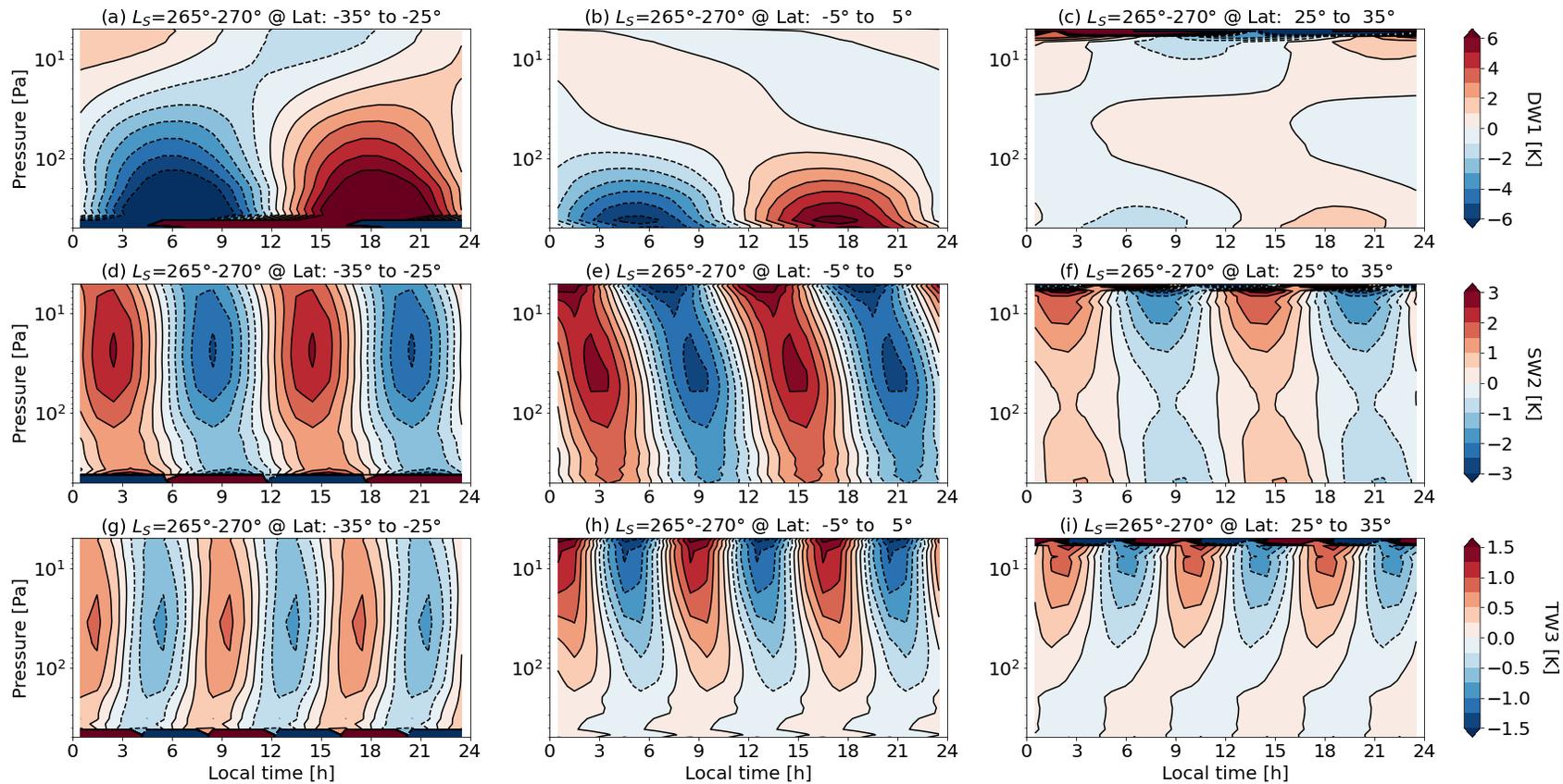

**Figure 8.** Same as Figure 6, but for the decomposition results using sampled and vertically convolved Mars PCM outputs (T$_C$).

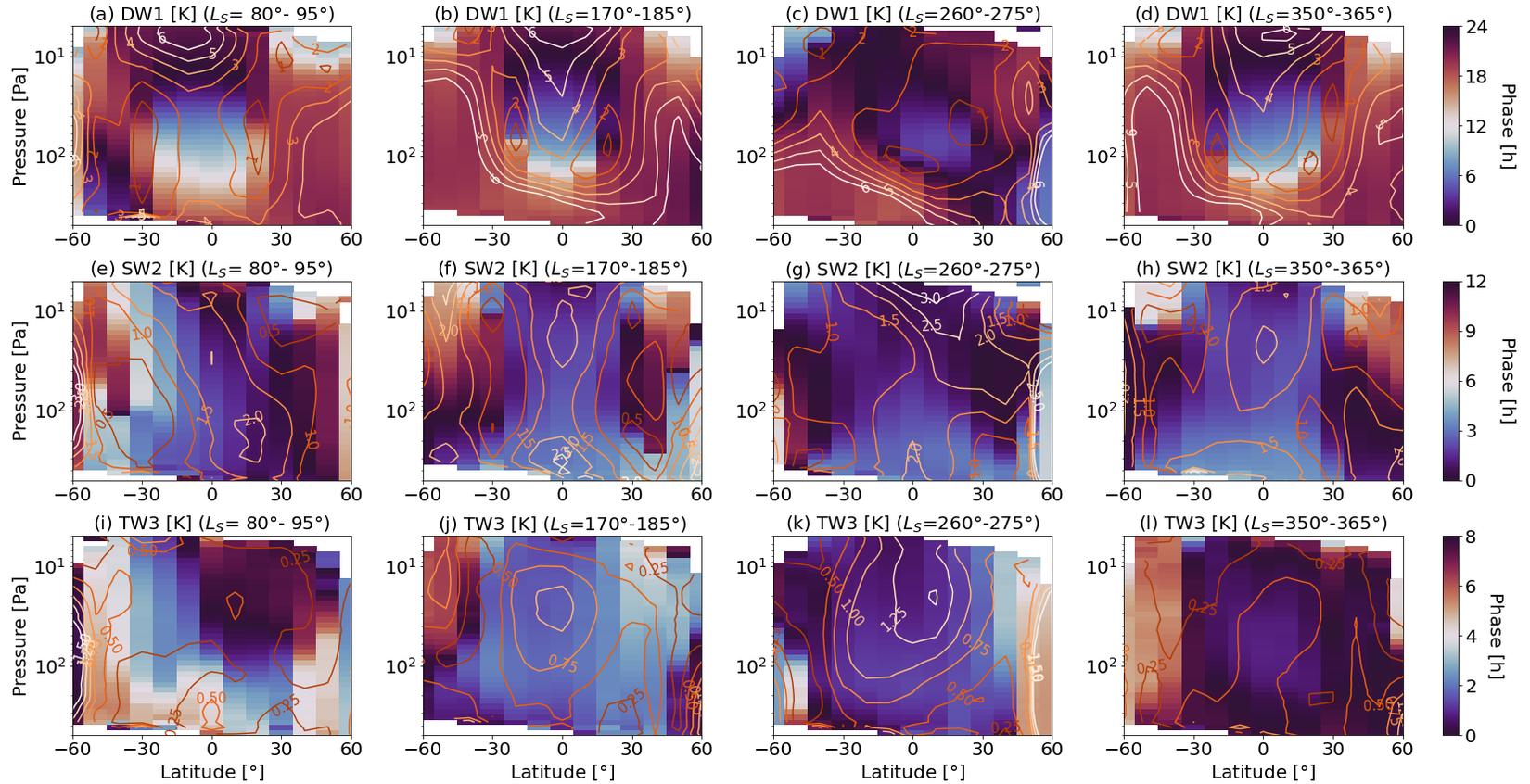

**Figure 9.** (a) Amplitude (colored contours) and Phase (colors) of the diurnal tide (DW1) derived through the wave mode decomposition using EMIRS observations (T) within the $L_S$ bin at MY 36 $L_S = 85° – 90°$ near the northern summer solstice. The colored contours have an interval of 1 K. The phase shown here is defined as the local time of the temperature maximum (wave crest). Only results with amplitude uncertainties of <5 K are shown. (b-d) Same as (a), but for the $L_S$ bin at MY 36 $L_S = 175° – 180°$, $265° – 270°$, and $355° – 360°$, respectively. (e-h) Same as (a-d), respectively, but for the semi-diurnal tide (SW2). The colored contours have intervals of 0.5 K, and the amplitude uncertainty threshold is 2.5 K. (i-l) Same as (a-d), respectively, but for the ter-diurnal tide (TW3). The colored contours have intervals of 0.25 K, and the amplitude uncertainty threshold is 1.25 K.

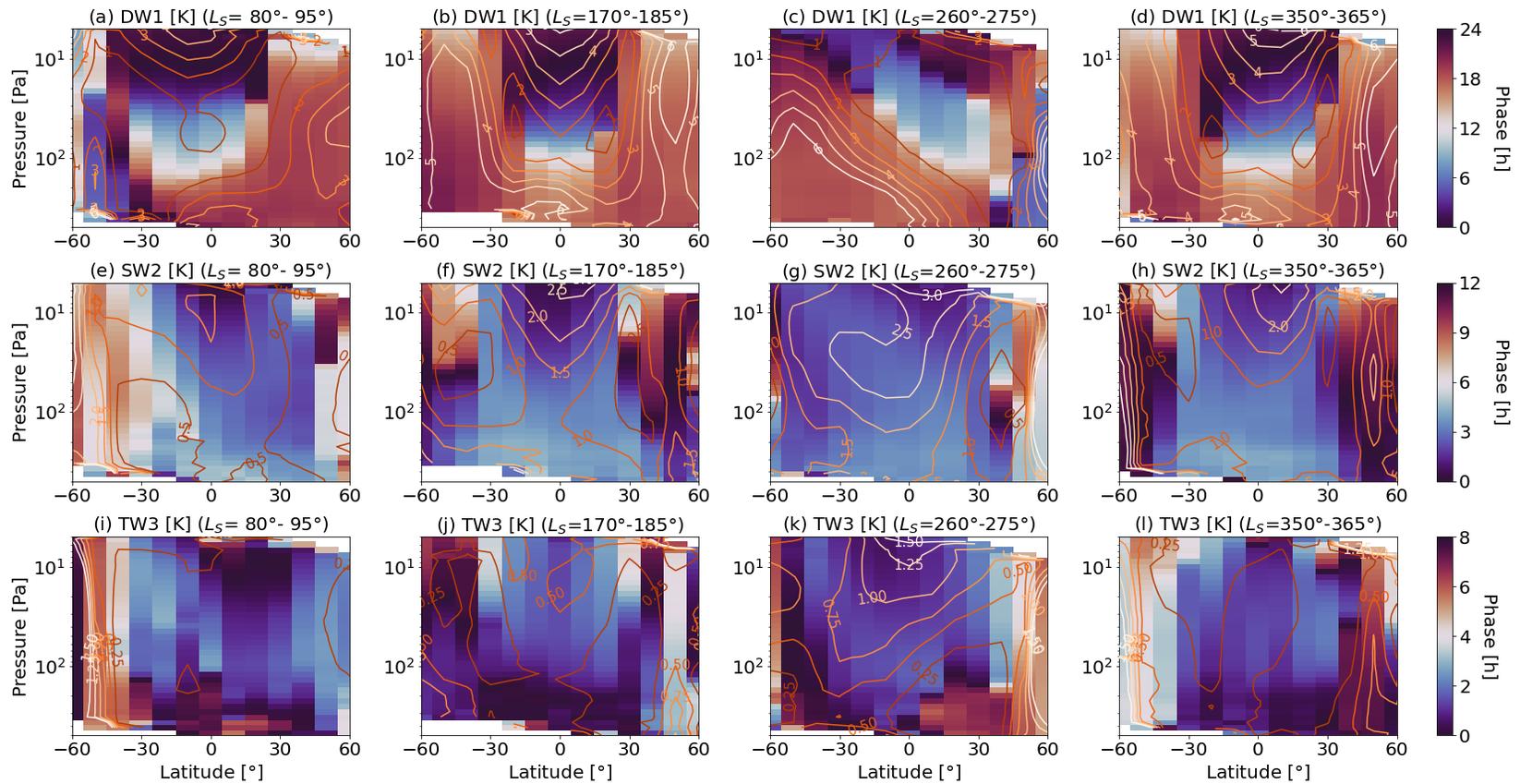

**Figure 10.** Same as Figure 9, but for the decomposition results using sampled and vertically convolved Mars PCM outputs ($T_C$).

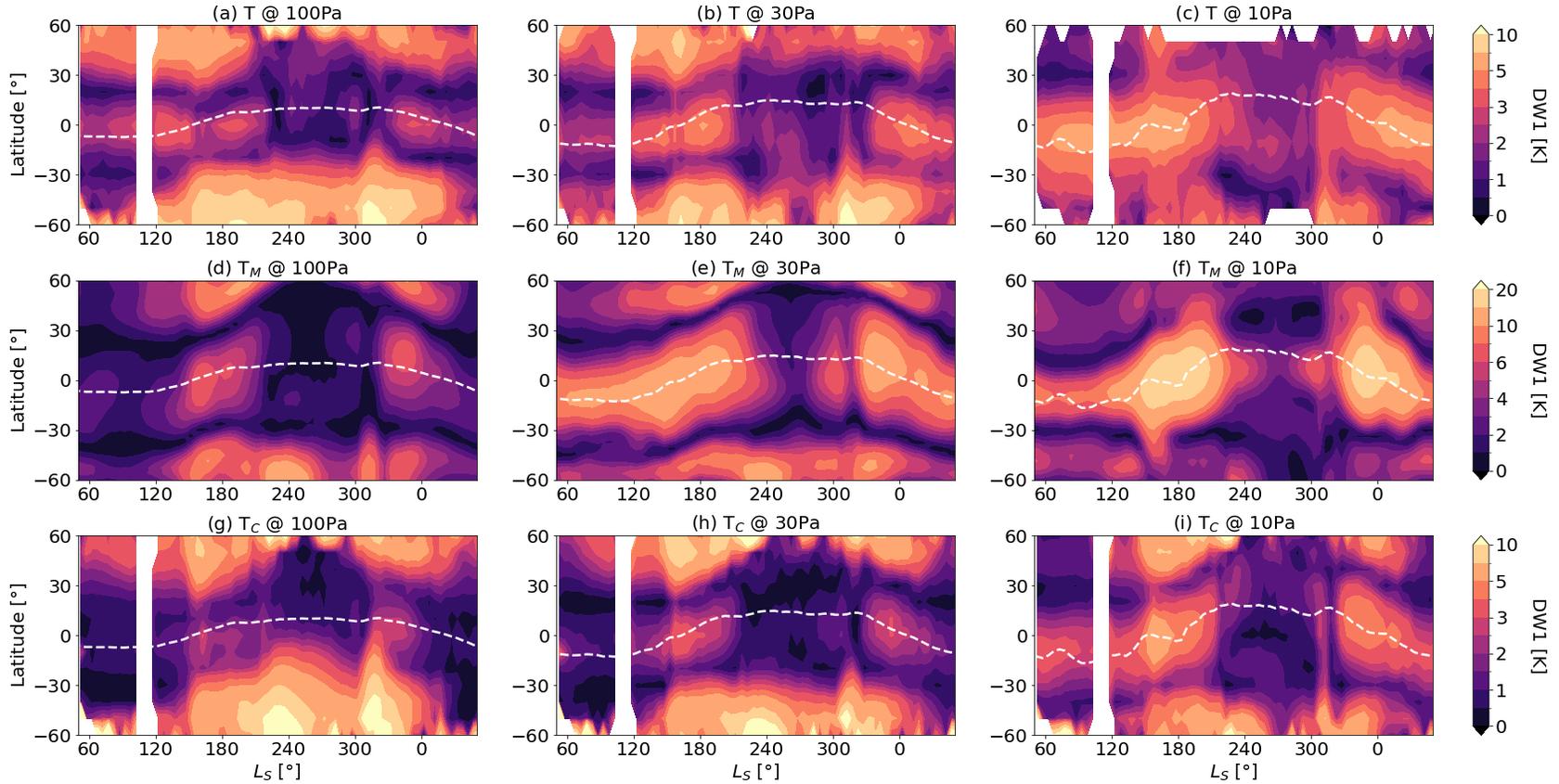

**Figure 11.** (a) Seasonal variation of the diurnal tide (DW1) amplitude at 100 Pa derived using EMIRS observations (T). The results presented for each 5° $L_S$ bin are analyzed using observations within the 15° $L_S$ range to improve the statistics, including the 5° $L_S$ bins immediately before and after. The white dashed line denotes the dynamical equator, defined as the latitude with zero zonal mean total vorticity, which is the sum of the Coriolis parameter and the vorticity of the zonal mean zonal winds simulated in the Mars PCM. The contours have non-uniform intervals for the purpose of presentation. Only results with amplitude uncertainties of <5 K are shown. (b, c) Same as (a), but for 30 Pa and 10 Pa, respectively. (d-f) Same as (a-c), but for the Mars PCM outputs ($T_M$). (g-i) Same as (a-c), but for the sampled and vertically convolved Mars PCM outputs ($T_C$).

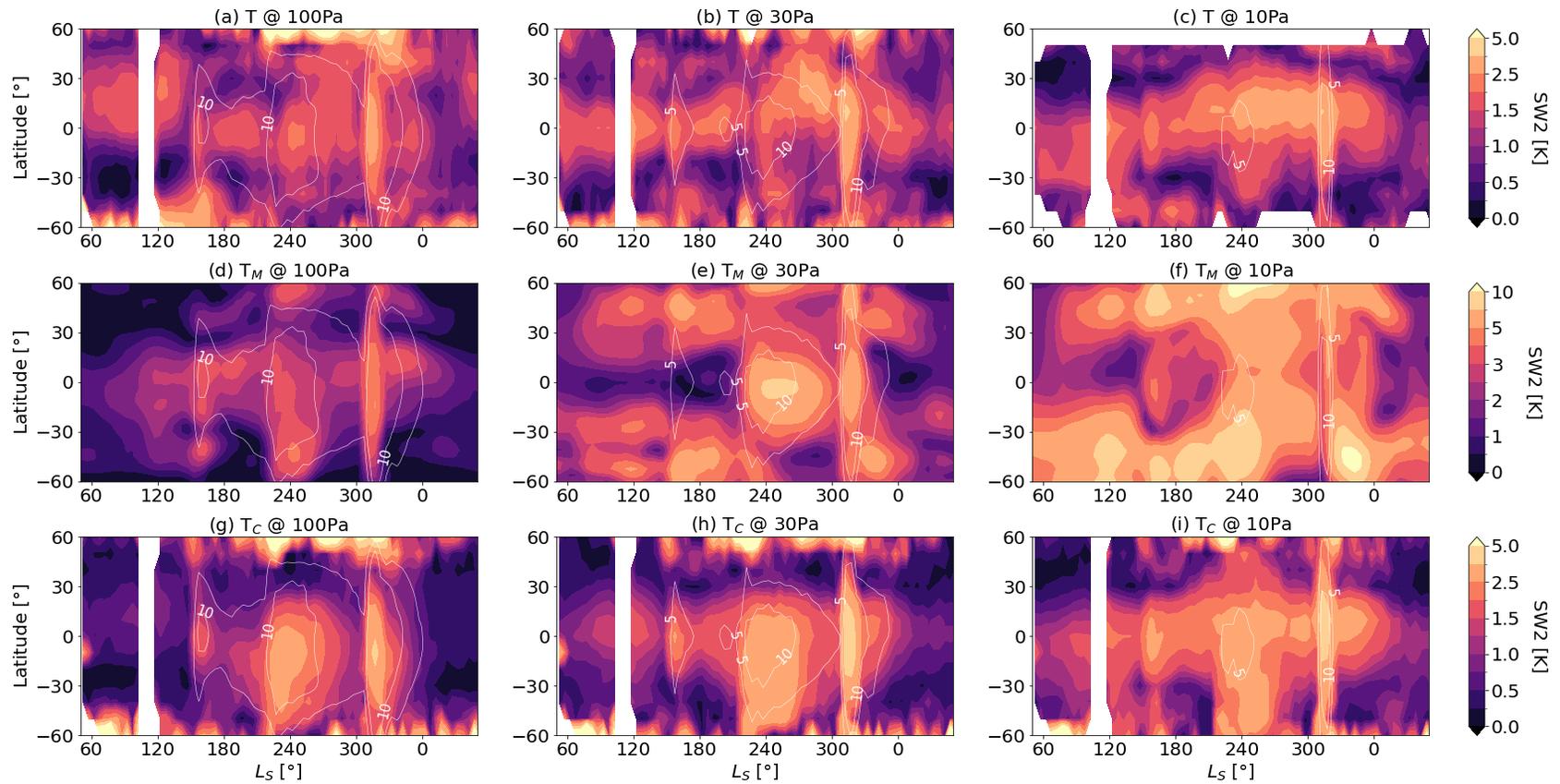

**Figure 12.** (a-i) Same as those in Figure 11, but for the semi-diurnal tide (SW2). The white contours denote the dust mass mixing ratios in the Mars PCM, which are multiplied by $10^6$ for the purpose of presentation. Only results with amplitude uncertainties of <2.5 K are shown.

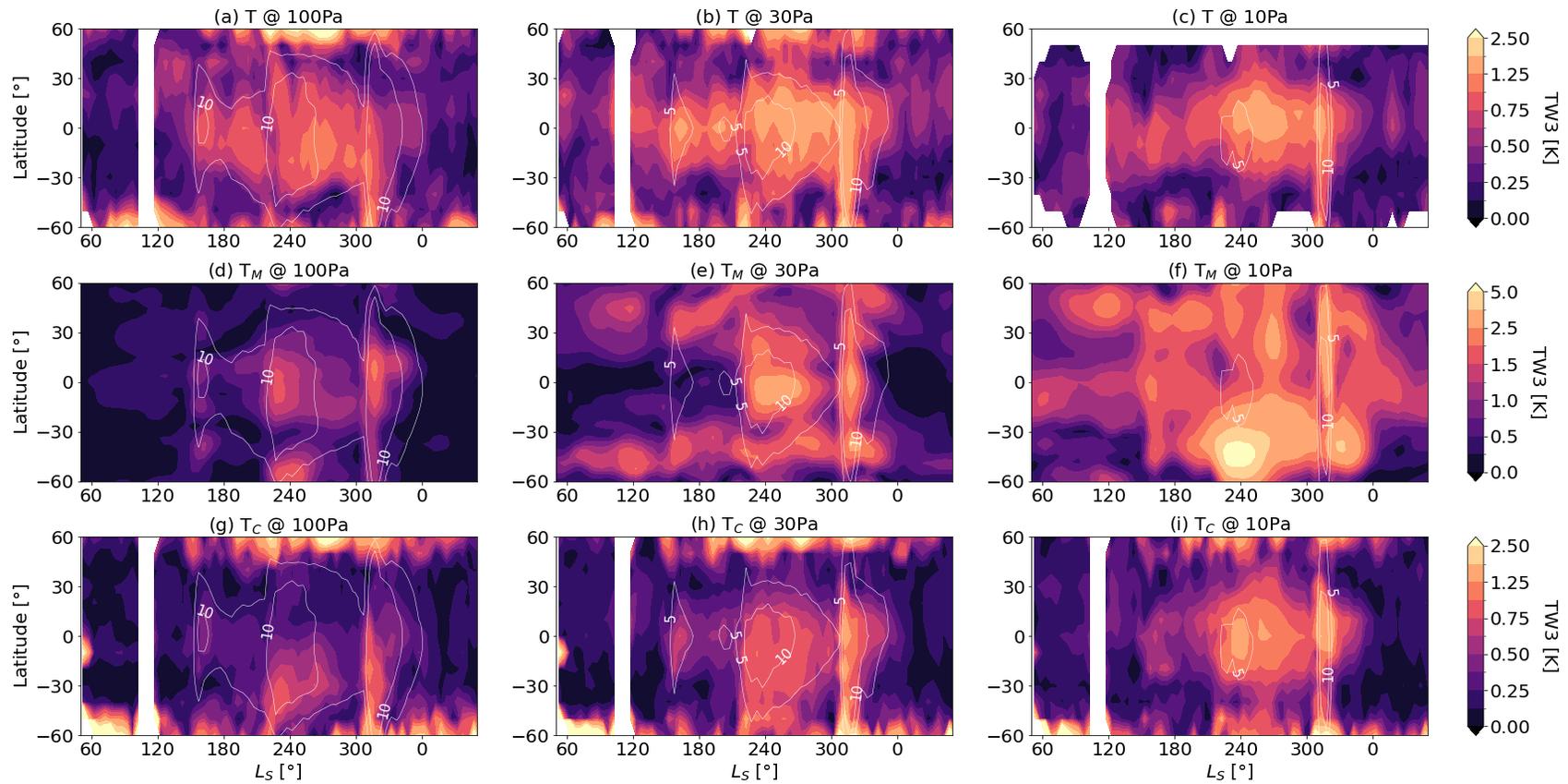

**Figure 13.** Same as Figure 12, but for the ter-diurnal tide (TW3). Only results with amplitude uncertainties of <1.25 K are shown.

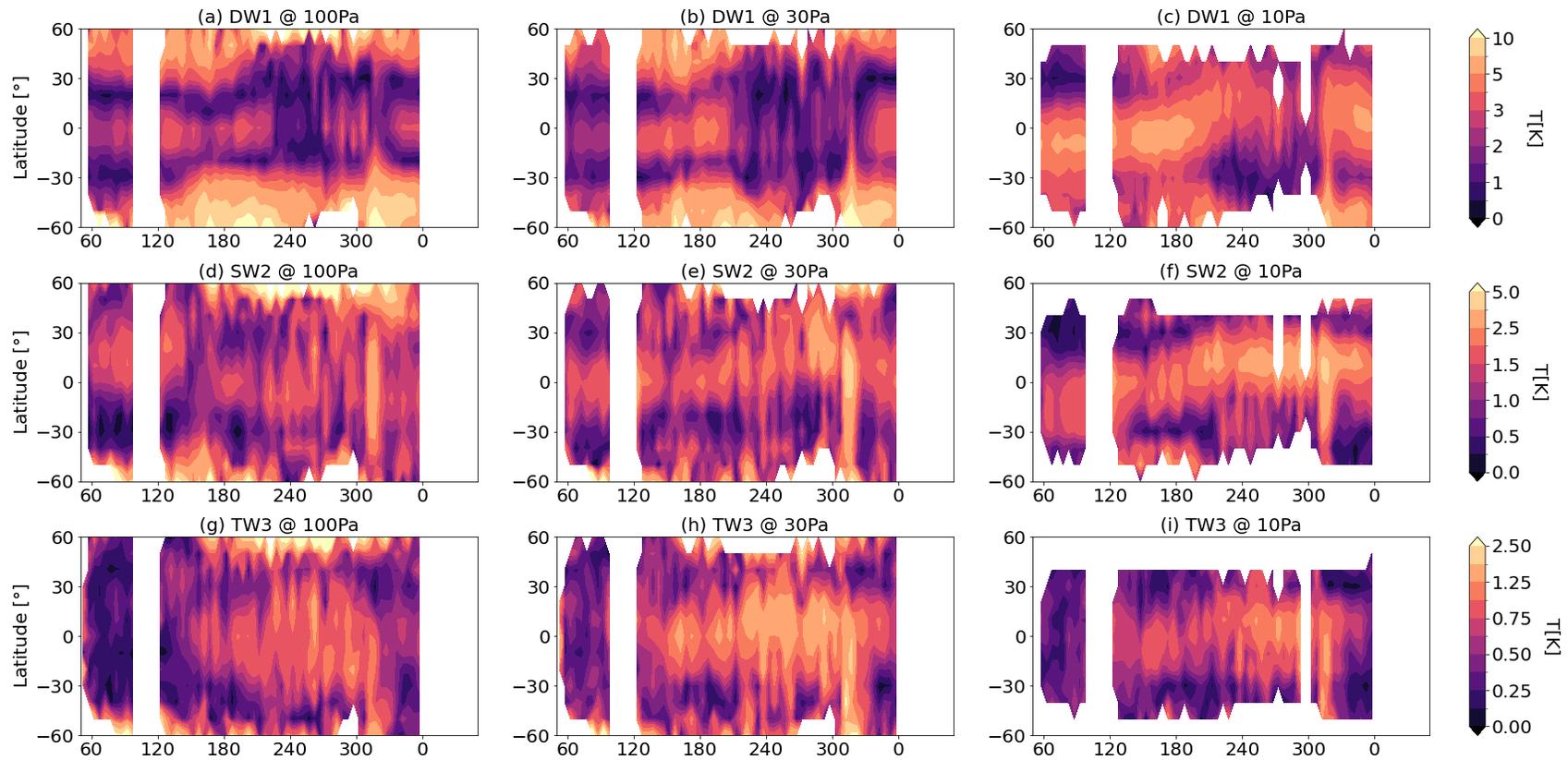

**Figure 14.** (a-c) Same as Figure 11a-11c, but for the results if only using the data within each 5° $L_S$ bin, not including any data before or after the $L_S$ range. (d-f) Same as (a-c), but for the semi-diurnal tide (SW2). (g-i) Same as (a-c), but for the ter-diurnal tide (TW3).